\documentclass[12pt,draftclsnofoot,onecolumn]{IEEEtran}

\usepackage{url}
\usepackage{graphicx}
\usepackage{balance}
\usepackage{graphics}
\usepackage{graphicx}
\usepackage{epsfig,amssymb}
\usepackage{epstopdf}
\usepackage{times}
\usepackage{mathrsfs}
\usepackage{array}
\usepackage{amsmath, latexsym, amsfonts, amssymb,bm}
\usepackage[mathscr]{eucal}
\usepackage{array, tabularx}
\usepackage[table]{xcolor}
\usepackage{multirow}
\usepackage{epsfig}
\usepackage{cite}
\usepackage{tikz}
\usepackage{mathtools}
\usepackage{psfragx}
\usepackage{xcolor}
\usepackage{bbm}
\usepackage{lineno}

\usepackage{algorithm}
\usepackage{algorithmic}

\usepackage{lipsum}

\usepackage{caption}
\usepackage{subcaption}
\usepackage{authblk}

\newcommand{\quotes}[1]{``#1''}

\newcommand{\defeq}{\ensuremath{\triangleq}} 
\newcolumntype{S}{>{\centering\arraybackslash}p{5em}}

\newtheorem{proposition}{Proposition}

\DeclareMathOperator{\Tr}{Tr}
\DeclareMathOperator{\rank}{rank}
\DeclareMathOperator{\diag}{{diag}}

\IEEEoverridecommandlockouts

\begin{document}
	
	\title{Intelligent Reflecting Surface: Practical Phase Shift Model and  Beamforming Optimization}
	
		\author{ 
			Samith Abeywickrama,   Rui Zhang,  \IEEEmembership{Fellow, IEEE}, Qingqing Wu, \IEEEmembership{Member, IEEE}, and  Chau Yuen,  \IEEEmembership{Senior Member, IEEE} 
			
			\thanks{
				Part of this work will be presented in the  IEEE International Conference on Communications, Dublin, Ireland, 2020 \cite{my_icc}.
				
				S. Abeywickrama is with the Department of Electrical and Computer Engineering, National University of Singapore, Singapore, and also with Singapore University of Technology and Design, Singapore (e-mail: samith@u.nus.edu).
				
				R. Zhang and Q. Wu are with the Department of Electrical and Computer Engineering, National University of Singapore, Singapore (e-mail: \{elezhang,elewuqq\}@nus.edu.sg).
				
				C. Yuen is with  Singapore University of Technology
				and Design, Singapore (e-mail: yuenchau@sutd.edu.sg).
			}
		}
	
%
	
	\maketitle 
	\vspace{-1.5cm}
	
	\begin{abstract}
		Intelligent reflecting surface (IRS) that enables the control of  wireless propagation environment has recently emerged  as a promising cost-effective  technology  for boosting  the spectrum and energy efficiency in future wireless communication systems. Prior works on IRS are mainly based on the \textit{ideal } phase shift model  assuming the  full signal  reflection by each of the elements regardless of its  phase shift, which, however, is practically difficult to realize. In contrast, we propose  in this paper the  \textit{practical  } phase shift model that captures the \textit{phase-dependent } amplitude variation  in the element-wise reflection coefficient. Based on the proposed model and considering  an IRS-aided multiuser  system with  an IRS  deployed to assist in the downlink  communications from  a multi-antenna access point (AP) to multiple single-antenna users, we formulate an optimization  problem to minimize the total transmit power at the AP by jointly designing  the AP transmit beamforming and the IRS reflect beamforming, subject to the users' individual signal-to-interference-plus-noise ratio (SINR) 		constraints. Iterative algorithms are  proposed to find  suboptimal solutions to this problem efficiently  by utilizing the  alternating optimization (AO) or penalty-based optimization technique. Moreover, we analyze the \textit{asymptotic performance loss}  of the  IRS-aided   system that employs practical phase shifters but  assumes the ideal phase shift model for  beamforming optimization, as the number of IRS elements goes to infinity.  Simulation results unveil  substantial performance gains  achieved by the proposed  beamforming optimization  based on the practical  phase shift model as compared to  the conventional ideal model.
		 
	\end{abstract}


	\IEEEpeerreviewmaketitle
	
	\section{Introduction}

	Intelligent reflecting surface (IRS) assisted wireless communication has recently emerged as a promising solution to enhance the spectrum  and energy efficiency of  future wireless systems cost-effectively. Specifically, an IRS is able to establish  favourable channel responses by controlling the wireless propagation environment  via  a large number of  reconfigurable passive reflecting elements (see e.g. \cite{wu2018intelligent,qq_magazine,chongwang,8796365,chongwen_hollo} and the references therein). 	In particular, it has been shown in \cite{wu2018intelligent} that when the number of reflecting elements, say  $N$, is  sufficiently large, IRS is able to achieve an asymptotic receive signal power or signal-to-noise ratio (SNR) gain of order $\mathcal{O}(N^2)$ in the IRS-aided single-user system, known as the \textit{squared power gain}. Moreover, for the multiuser system,  by jointly optimizing the active (transmit) beamforming at the base station  and the passive (reflect) beamforming at the IRS, the signal-to-interference-plus-noise ratio (SINR) performance of all users in the network can be significantly improved  \cite{wu2018intelligent}, regardless of whether they are aided  by the IRS directly or not. The proposed joint active and passive beamforming design has  been also investigated in  various other applications/setups, e.g., physical layer security \cite{8723525,8972400,xu2019resource}, orthogonal frequency division multiplexing (OFDM)  \cite{irs_ofdm,IRS_channel_estimation},  non-orthogonal multiple access (NOMA) \cite{yang2019intelligent,8970580}, and simultaneous wireless information and power transfer (SWIPT)  \cite{8941080,IRS_SWIPT_2}. However, the above prior  works as well as many others (see, e.g.,  \cite{nadeem2019intelligent,wang2019joint,zhang2019capacity,pan2019multicell,fu2019intelligent,mu2019exploiting,zhao2019intelligent})  on IRS have all assumed the  ideal phase shift model with full signal  reflection, i.e., unity reflection  amplitude at each reflecting  element regardless of its phase shift, which, however, is practically difficult to realize due to the hardware limitation \cite{4619755,phase_dependent_amplitude}.
	
	The \textit{amplitude response}  of a typical passive reflecting element is generally non-uniform with respect to its phase shift. In particular, the amplitude typically  exhibits its minimum value at the zero phase shift, but monotonically increases and asymptotically approaches  unity amplitude (i.e., $1$) at the phase shift of $\pi$ or $-\pi$. This is due to the fact that when the phase shift approaches zero, the image currents, i.e., the currents of a virtual source that accounts for the reflection, are in-phase with the reflecting element currents, and thus the electric field and the current flow in each  element are  enhanced. As a result, the dielectric loss, metallic loss, and ohmic loss  increase accordingly, leading to  more energy loss  and hence lower reflection amplitude \cite{4619755}. Furthermore, these losses mainly come from the semiconductor devices, metals, and dielectric substrates used in  IRS,  thus cannot be completely avoided  in practice. In fact,  this is a long-standing problem  for designing  reflection-based metasurfaces \cite{zhu2013active}.  In \cite{irs_amp}, amplifiers are integrated into the reflecting elements to compensate the energy loss, which, however, is unsuitable  for passive IRS and also costly to implement in practice.
	
	In prior  works (e.g., \cite{8723525,8972400,xu2019resource,irs_ofdm,IRS_channel_estimation,yang2019intelligent,8970580,8941080,IRS_SWIPT_2,nadeem2019intelligent,wang2019joint,zhang2019capacity,pan2019multicell,fu2019intelligent,mu2019exploiting,zhao2019intelligent})  on IRS, by assuming the ideal phase shift model with unity  reflection amplitude regardless of the phase shift at each element,  IRS reflection has been  designed to achieve  the maximal  phase alignment between the IRS-reflected and non-IRS-reflected signals at the designated receiver(s).
	However, when the reflection amplitude depends on the phase shift, such a reflection design is  no longer optimal in general and will cause  performance degradation. Instead, the phase shifts at the IRS need to be properly designed  to strike  an optimal  balance between the reflected signal  amplitude and phase alignment under the practical phase shift model with phase-dependent amplitude response,  which thus  motivates this work.
	
	
	
\begin{figure}[t!]
		\centering \vspace{-0.8cm}
		\includegraphics[width=0.5\linewidth]{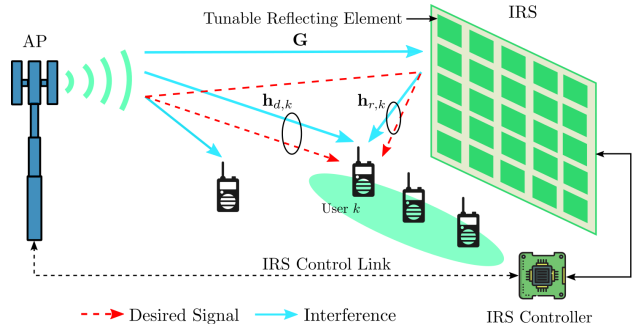}
		\caption{{An  IRS-aided multiuser wireless  communication system.}}  
		\label{system_model} \vspace{-0.5cm}
	\end{figure}

	In this paper, by considering the  practical IRS phase shift model and  an IRS-aided multiuser wireless communication system  shown  in Fig. \ref{system_model},  we formulate and solve new problems to minimize the total transmit power at the access point
	(AP) by jointly optimizing the AP transmit beamforming and the IRS reflect beamforming, subject to the users' individual SINR		constraints. Our main contributions are summarized as follows.
		
	
	\begin{itemize} 
		\item First,  to characterize  the fundamental  relationship between the reflection  amplitude and  phase shift for designing IRS-aided wireless systems, we propose  a new  \textit{analytical model}
		for the phase shifter, which is   applicable to a variety of semiconductor devices used for implementing  IRS, and verify its accuracy using  the experimental results from  the literature.
		\item Next, based on the newly established  phase shift model, we solve the formulated problem for  the special case of single-user transmission to draw useful design insights. However, the problem in this case  is non-convex and difficult to be optimally solved in general, for which we propose two iterative algorithms to find  suboptimal solutions efficiently by utilizing the alternating optimization (AO) and penalty-based optimization techniques, respectively. Moreover,  to highlight the importance of the practical phase shift model, we analyze the asymptotic performance loss  of an IRS-aided   system that employs practical phase shifters but assumes the ideal phase shift model for  beamforming optimization, as the number of IRS elements becomes large.
		\item Finally,  we extend our formulated problem   for the single-user case to the more general multiuser case and propose two algorithms to obtain suboptimal solutions for it,  which  offer different tradeoffs between the  complexity and performance. Specifically, the first algorithm, which is based on the  penalty-based method, achieves better performance but requires   higher complexity as compared to the second algorithm, which is designed by following the similar  principle of  two-stage optimization as proposed in \cite{wu2018intelligent}. 		
		Simulation results unveil substantial performance gains  achieved by the proposed  beamforming optimization  based on the practical  phase shift model as compared to  the conventional ideal model, although the latter  has been widely adopted  in the literature. 
		
	\end{itemize}

	The rest of this paper is organized as follows. Section \ref{system_model_sec} 	introduces the system model. Section \ref{phase_shift_model_sec} presents the proposed practical IRS phase shift model. Section \ref{su} presents the problem formulation for  the single-user case and proposes two efficient algorithms to solve the  problem. In Section \ref{mu_sec}, we extend  the algorithms proposed for the single-user case to the  general multiuser case. Section \ref{sr_sec} presents numerical results to evaluate the effectiveness  of the proposed algorithms. Finally, we conclude the paper in Section \ref{con_sec}.

	\textit{Notations:} In this paper, scalars are denoted by italic letters, vectors and matrices are denoted by bold-face lower-case and upper-case letters, respectively. For a complex-valued vector $ \mathbf v $, $ \| \mathbf v \| $, $ \mathbf v^H $,  and $\diag(\mathbf v)$ denote its $\ell_2$-norm, conjugate transpose, and a diagonal matrix with each diagonal element being the corresponding element in $ \mathbf v $, respectively. Scalar $v_i$ denotes the $i$-th element of a vector $\mathbf v$. For a square matrix $ \mathbf{S} $, $ \Tr(\mathbf{S}) $ and $ \mathbf{S}^{-1} $ denote its trace and inverse, respectively, while $\mathbf{S} \succeq 0$  means that $ \mathbf{S} $ is positive semi-definite. For any  matrix $ \mathbf{A} $,  $ \mathbf{A}^H $, $ \rank(\mathbf{A}) $, and $\mathbf{A}_{n,k}$ denote its conjugate transpose, rank, and $(n, k)$th element, respectively. $ \mathbf{I} $ and $ \mathbf{0} $ denote an identity matrix and an all-zero matrix, respectively, with appropriate dimensions.
	$ \mathbb{C}^{x \times y} $  denotes the space of $ x \times y $ complex-valued matrices. $ \jmath $ denotes the imaginary unit, i.e., $ \jmath^2 = -1 $.  For a complex-valued scalar $  v $, $ |  v | $ and $\arg(v)$ denote its absolute value and phase, respectively. The distribution of a circularly symmetric complex Gaussian (CSCG) random vector with mean $\mu$ and variance $\varrho^2$ is denoted by $\mathcal{CN}(\mu,\,\varrho^{2})$; and $\sim$ stands for \quotes{distributed as}. $\mathbb{E}(\cdot)$ denotes the statistical expectation.
	
	
	\section{System Model} \label{system_model_sec}
	
	As shown in Fig. \ref{system_model}, we consider a multiuser  multiple-input single-output (MISO) wireless system, where an IRS composed of $N$ reflecting elements is deployed to assist in the downlink  communications from an AP with $M$ antennas to $K$ single-antenna users. 
	The set of the users is denoted by $\mathcal K=\{1,\dots,K\}$. 
	The IRS reflecting elements are  programmable via a smart  IRS controller \cite{qq_magazine}. Furthermore, the IRS controller communicates with the AP via a separate wireless link for the AP to control the IRS reflection. 
	It is  assumed that the signals that are reflected by the IRS more than once  have negligible power due to substantial path loss and  thus are ignored \cite{wu2018intelligent}. 
	In addition, we consider a quasi-static flat-fading model, where it is assumed that all the wireless channels remain constant over each transmission block. All the  channels are assumed to be known at the AP by applying, e.g., the channel estimation techniques proposed in \cite{qq_magazine,IRS_channel_estimation}.  
	
	Let $\mathbf h_{d,k} \in \mathbb C^{M\times1}$, $\mathbf h_{r,k} \in \mathbb C^{N\times1}$, and $\mathbf G \in \mathbb C^{N\times M}$ denote the  baseband equivalent channels from the AP to user $k$, from the IRS to user $k$, and from the AP to IRS, respectively,  $ k\in \mathcal K$. Without loss of generality, let  $\mathbf v=[v_1,\dots,v_N]  \in \mathbb C^{N\times1}$   denote the reflection coefficient vector of the IRS, where $|v_n| \in [0,  1]$ and $ \arg(v_n) \in [-\pi,  \pi)$  are the reflection amplitude and  phase shift on the combined incident signal, respectively, for $n \in \{1,\dots,N\} $ \cite{qq_magazine}. Note that for the ideal phase shift model considered in \cite{wu2018intelligent,qq_magazine,chongwang}, it follows that $|v_n|=1,\forall n$, regardless of the phase shift, $\arg(v_n)$. The  transmit signal at the AP is given by $ \mathbf{x} = \sum_{k=1}^{K} \mathbf{w}_k s_k $, where $\mathbf w_k \in \mathbb{C}^{M \times 1}$ denotes the transmit  beamforming vector for user $k$ and $ s_k $ denotes the corresponding transmit symbol, which is independent over $k$, and has zero-mean and unit-variance ({i.e.,} $ \mathbb{E}( |s_k|^{2}) =1 $). The signal
	received at user $k$ from both the AP-user and AP-IRS-user channels is then expressed as 
	\begin{align} \label{received_signal}
	y_k &=  (\mathbf v^H  \mathbf \Phi_k + \mathbf h_{d,k}^H)\sum_{j=1}^{K} \mathbf{w}_j s_j + z_k,  k \in \mathcal K,
	\end{align} 
	where $\mathbf \Phi_k=   \diag(\mathbf h_{r,k}^H) \mathbf G$ and $z_k$ denotes the additive white Gaussian noise (AWGN) at the user $k$'s
	receiver with zero-mean and variance $\sigma^2_k$. Accordingly, the SINR of user $k$ is given by
	\begin{align} \label{r_se} 
	\text{SINR}_k &=  \frac{|(\mathbf v^H  \mathbf \Phi_k + \mathbf h_{d,k}^H)\mathbf{w}_k|^2}{\sum_{j\neq k}^{K} |(\mathbf v^H  \mathbf \Phi_k + \mathbf h_{d,k}^H)\mathbf{w}_j|^2 + \sigma^2_k},  k\in \mathcal K.
	\end{align} 
		
	\section{Practical  Phase Shift Model} \label{phase_shift_model_sec}
	
	\begin{figure}[t!]
		\centering \vspace{-0.8cm}
		\includegraphics[trim = 0mm 0mm 0mm 0mm, clip,width=0.5\linewidth]{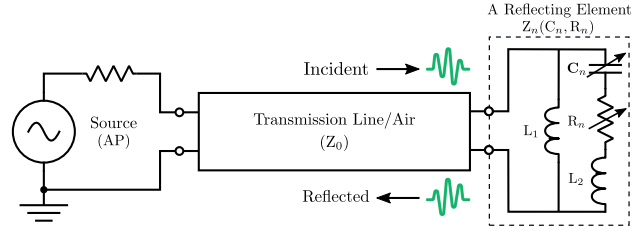}
		\caption{{Transmission line model of a  reflecting element.}}  
		\label{transmission_model} \vspace{-0.5cm}
	\end{figure}

	\subsection{Equivalent Circuit Model} \label{circuit_sec}

	An IRS is typically constructed by using  the  printed circuit board (PCB), where the reflecting elements are equally spaced  on a two-dimensional plane. A  reflecting element is composed of a metal patch on the top layer of the PCB dielectric substrate and a full metal sheet on the bottom layer \cite{qq_magazine}. 
	Moreover, a semiconductor device\footnote{In practice, a positive-intrinsic-negative (PIN) diode, a variable capacitance (varactor) diode, or a metal-oxide-semiconductor field-effect transistor (MOSFET) can be used as the semiconductor device mentioned here  \cite{tang2018wireless,zhu2013active,PhysRevApplied}. }, which can vary the impedance of the reflecting element by controlling its biasing voltage, is embedded into the top layer metal patch  so that the element response  can be dynamically tuned in real time without changing the geometrical parameters \cite{PhysRevApplied}. In other words, when the geometrical parameters are fixed, the semiconductor device  controls the phase shift and reflection  amplitude (absorption level).
	
	As the physical length of a reflecting element is usually smaller than  the wavelength of the desired incident  signal, its response can be accurately
	described by an equivalent lumped circuit model regardless of the particular
	geometry of the element \cite{koziel2013surrogate}.
	As such, the metallic parts in the reflecting element can be modeled as inductors as the high-frequency current flowing in it produces a quasi-static magnetic field. In  Fig. \ref{transmission_model}, the equivalent model for the $n$-th reflecting element is illustrated  as a parallel resonant circuit and its impedance is given by
	\begin{align} \label{z_n}
	Z_n(C_n,R_n) &=  \frac{\jmath \omega L_1 (\jmath\omega L_2+\frac{1}{\jmath \omega C_n}+R_n) }{\jmath \omega L_1 + (\jmath\omega L_2+\frac{1}{\jmath \omega C_n}+R_n)},
	\end{align}
	where $L_1$, $L_2$, $C_n$, $R_n$, and $\omega$ denote the bottom layer inductance, top layer inductance, effective capacitance, effective resistance, and angular frequency of the incident signal, respectively. 
	Note that $R_n$ determines the amount of power dissipation due to the losses in the semiconductor devices, metals, and dielectrics, which cannot be zero in practice, and $C_n$ specifies  the charge accumulation related to  the element geometry and semiconductor device. 	As the transmission line diagram in Fig. \ref{transmission_model} depicts, the reflection coefficient, i.e., $v_n$ in \eqref{received_signal}, is the parameter that describes the fraction of the reflected electromagnetic wave  due to the impedance discontinuity between the free space impedance  $Z_0$ and element impedance $Z_n(C_n,R_n)$  \cite{microwave_book}, which  is given by  
	\begin{align} \label{v_n}
	v_n &=  \frac{Z_n(C_n,R_n) - Z_0}{Z_n(C_n,R_n) + Z_0}.
	\end{align}
	Since $v_n$ is a function of $C_n$ and $R_n$, the reflected electromagnetic waves can be manipulated  in a controllable and programmable manner by varying $C_n$'s and $R_n$'s. 
	
	\begin{figure} 
		\centering \vspace{-0.8cm}
		\begin{subfigure}[b]{0.45\textwidth} 
			\centering
			\includegraphics[trim = 0mm 0mm 0mm 0mm, clip,width=\columnwidth]{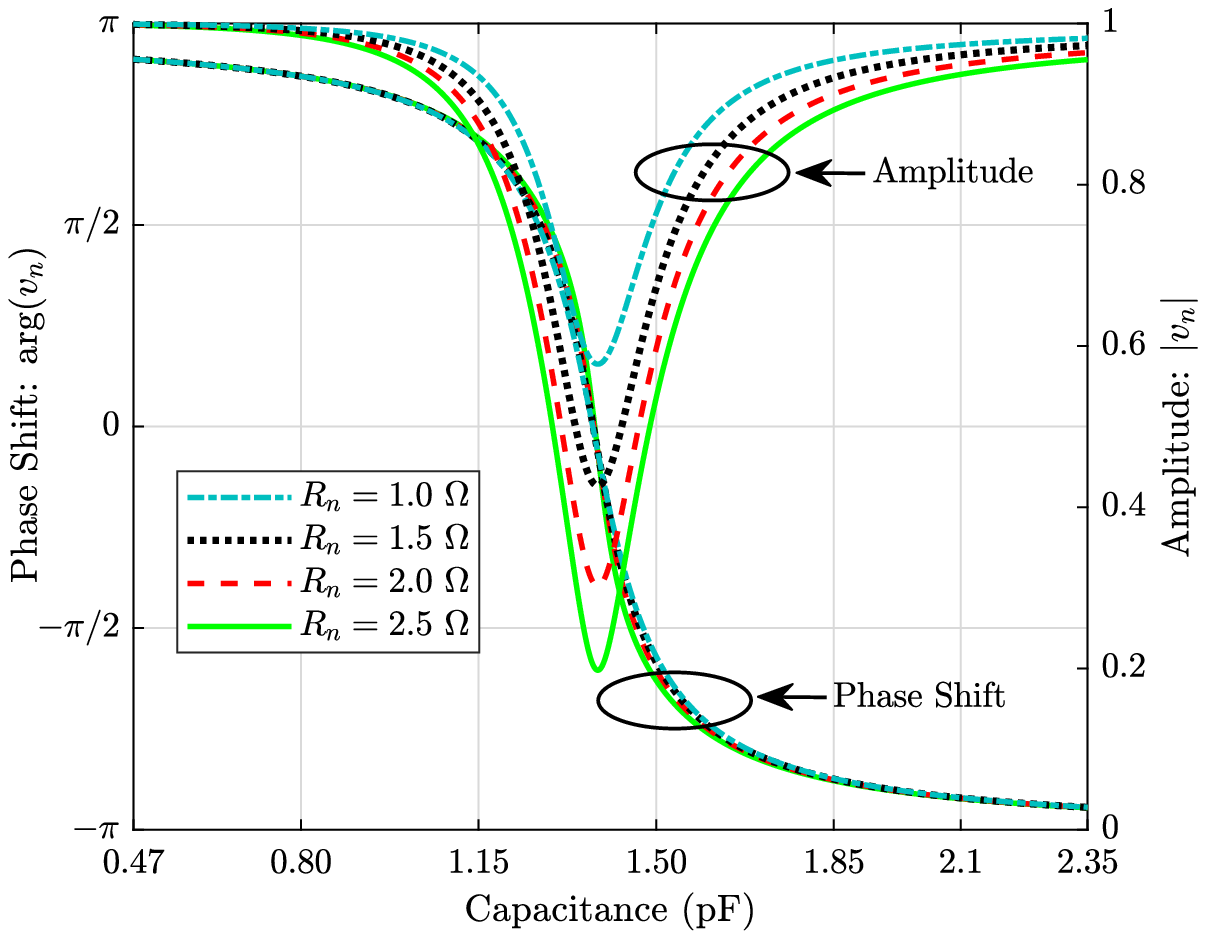}
			\caption{Phase shift and amplitude  versus  $C_n$ and $R_n$.}
			\label{fig_3_a} 
		\end{subfigure}
		\hspace{-0.1cm}
		\begin{subfigure}[b]{0.45\textwidth} 
			\centering
			\includegraphics[trim = 0mm 0mm 0mm 0mm, clip,width=\columnwidth]{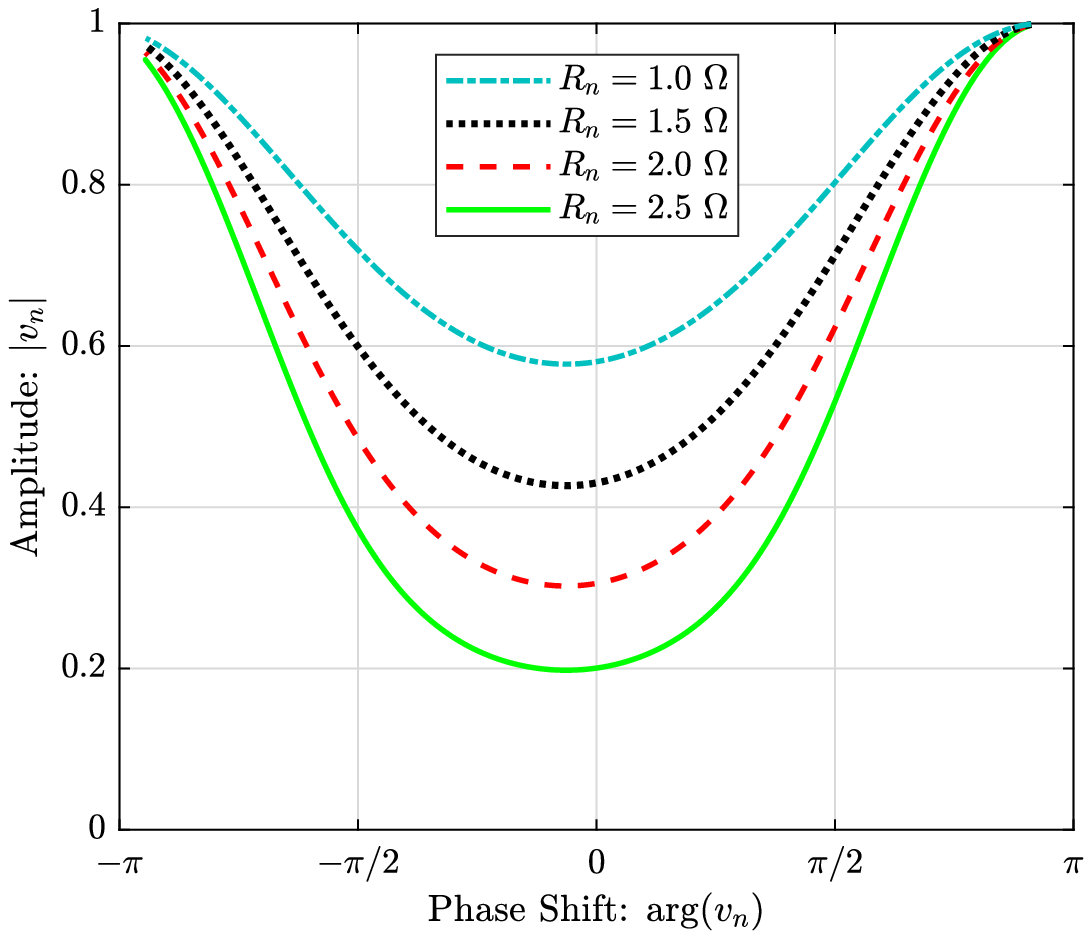}
			\caption{Amplitude versus  phase shift. }
			\label{fig_3_b} 
		\end{subfigure}
		\caption{Reflection coefficient of a  reflecting element.} \label{reflection} \vspace{-0.8cm}
	\end{figure} 
	
	To demonstrate this, Fig. \ref{reflection} illustrates the behaviour of the amplitude  and the phase shift, i.e., $ |v_n|$ and $\arg(v_n)$, respectively, for different values of $C_n$ and $R_n$.  Note that to align with the experimental results in \cite{zhu2013active}, $C_n$ is varied from $0.47$ pF to $2.35$ pF when $L_1=2.5$ nH, $L_2=0.7$ nH, $Z_0=377 \ \ \Omega$, and $\omega=2\pi \times 2.4 \times 10^9$. It is observed that although  a reflecting element is capable of achieving almost $2\pi$ full phase tuning, its phase shift and reflection  amplitude both  vary with  $C_n$ and $R_n$ in general. One can also observe  that the minimum amplitude occurs  near zero  phase shift and  approaches unity (the maximum value)  at the phase shift of $\pi$ or $-\pi$, which is explained as follows. When the phase shift is around $\pi$ or $-\pi$, the reflective currents (also termed as image currents) are out-of-phase with the element currents, and thus the electric field  as well as  the current flow in the element are both  diminished, thus resulting in minimum  energy loss and the  highest  reflection amplitude. In contrast, when the phase shift is around  zero, the reflective currents are in-phase with the element currents, and thus the electric field  as well as  the current flow in the element are both  enhanced. As a result, the dielectric loss, metallic loss, and ohmic loss increase dramatically, leading to  maximum  energy dissipation and the  lowest reflection amplitude. Furthermore, it is worth pointing out  that the numerical results illustrated in Fig. \ref{reflection} are in accordance with  the experimental results reported in the literature (see \cite{4619755} and Fig. 5 (b) in \cite{zhu2013active}), indicating that the circuit model given by \eqref{z_n} and \eqref{v_n}  accurately captures the physical reflection  of a reflecting element in practice.

	Note that to obtain the ideal phase shift control, where $|v_n| = 1,\forall \arg(v_n) \in [-\pi,  \pi)$, each  element should exhibit zero energy dissipation for reflection. However, in  practical hardware, energy dissipation is unavoidable\footnote{In \cite{zhu2013active}, $R_n=2.5$ $\Omega$ in each reflecting element due to the diode junction resistance, while in  \cite{4619755}, although the reflecting element does not contain any semiconductor device, its amplitude response follows a similar shape to  Fig. \ref{reflection} due to the metallic loss and dielectric loss.} and the typical behaviour of the reflection amplitude is similar to Fig. \ref{reflection}.   Therefore, incorporating the practical phase shift model to design  beamforming  algorithms  is essential to optimize  the performance of  IRS-aided wireless systems.
	
	\subsection{Proposed Phase Shift Model}
	
	\begin{figure} 
		\centering \vspace{-0.8cm}
		\begin{subfigure}[b]{0.42\textwidth} 
			\centering
			\includegraphics[trim = 0mm 0mm 0mm 0mm, clip,width=\columnwidth]{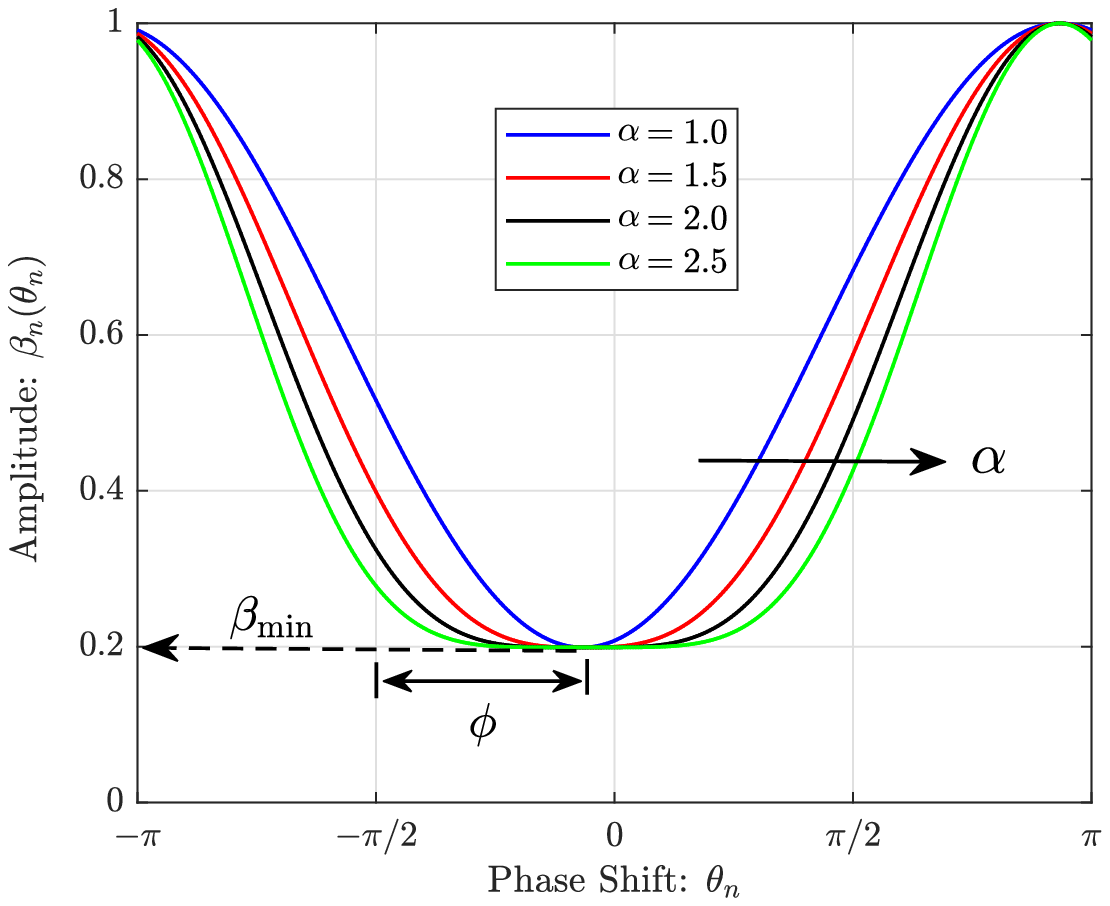}
			\caption{The phase shift model with different parameters.}
			\label{fig_4_a} 
		\end{subfigure}
		\hspace{1cm}
		\begin{subfigure}[b]{0.42\textwidth} 
			\centering
			\includegraphics[trim = 0mm 0mm 0mm 0mm, clip,width=\columnwidth]{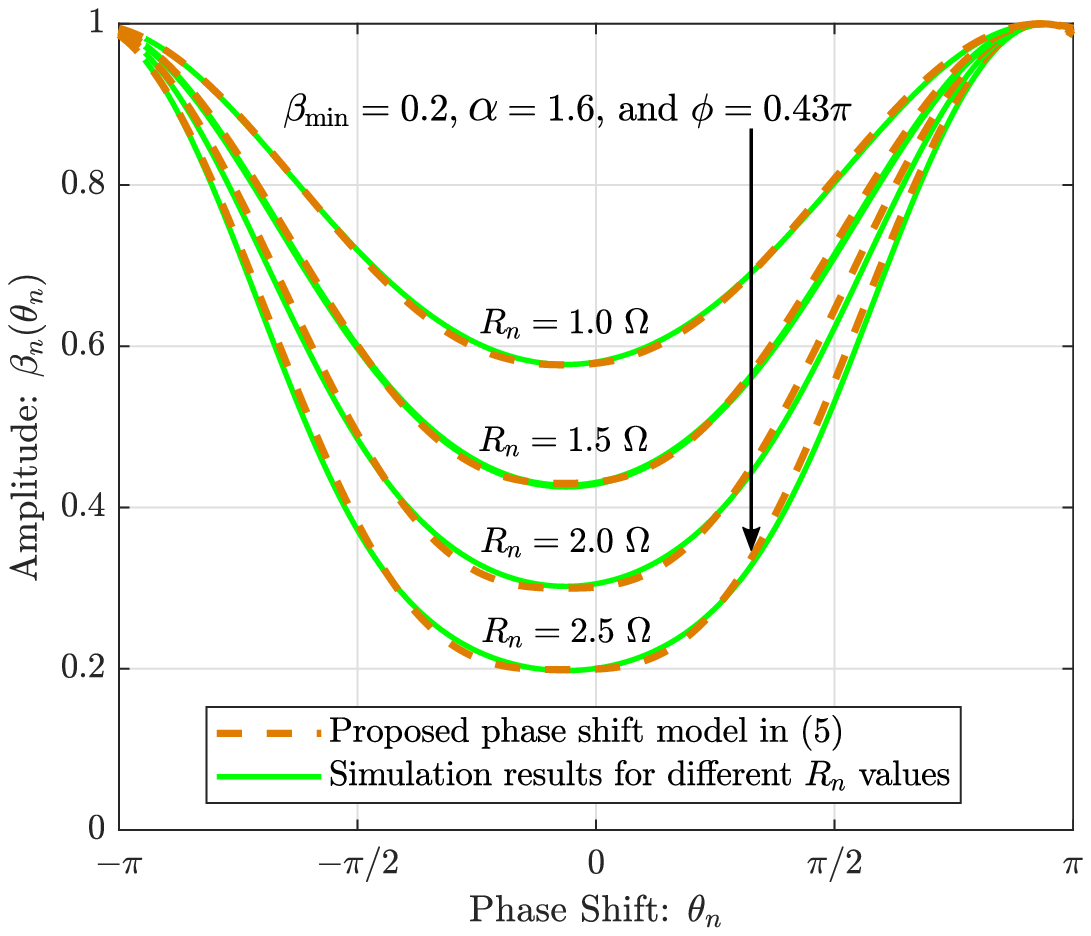}
			\caption{Simulation results for the proposed  phase shift model. }
			\label{fig_4_b} 
		\end{subfigure}
		\caption{The proposed phase shift model.} \label{model} \vspace{-0.8cm}
	\end{figure}
	
	In order to characterize  the fundamental  relationship between the reflection  amplitude and  phase shift for designing IRS-aided wireless systems, we propose in this subsection an analytical model for the phase shift which is in general   applicable to a variety of semiconductor devices used for implementing  IRS. Let $v_n = \beta_n(\theta_n) e^{\jmath \theta_n} $ with $\theta_n \in [-\pi,  \pi)$ and $\beta_n(\theta_n) \in [0,  1]$ respectively  denote  the phase shift and the corresponding amplitude. Specifically, $\beta_n(\theta_n)$ can be expressed  as 
	\begin{align} \label{our_model}
	\beta_n(\theta_n) &=  (1-\beta_{\text {min}}) \left( \frac{\sin(\theta_n - \phi) +1}{2} \right)^\alpha + \beta_{\text {min}},
	\end{align}
	where $\beta_{\text {min}}\geq 0$, $\phi \geq 0$, and $\alpha\geq0$ are the constants related to the specific circuit implementation. As depicted in Fig. \ref{model} (a), $\beta_{\text {min}}$ is the minimum amplitude, $\phi$ is the  horizontal distance between $-\pi/2$ and $\beta_{\text {min}}$,  and $\alpha$ controls the steepness of the function curve. Note that for $\beta_{\text {min}}=1$ (or $\alpha=0$), \eqref{our_model} is equivalent to the ideal phase shift model with unity amplitude. In practice, IRS  circuits are fixed once they are fabricated and thus these parameters can be easily found by a standard curve fitting tool.
	
	Fig. \ref{model} (b) illustrates that the proposed phase shift model  closely matches the simulation results presented in Section \ref{circuit_sec} for a practical reflecting element. In the sequel, we will adopt the model in \eqref{our_model}  for beamforming design in IRS-aided wireless communication. Without loss of generality,  we assume that the  circuits of the reflecting elements are all identical, and thus the same model parameters, i.e., $\beta_{\text {min}}$, $\phi$, and $\alpha$,  apply to  each of the elements at the IRS.

	\section{Single-User Beamforming Optimization} \label{su}
	
	In this section, we consider the special case with a single user to draw important insights into the   beamforming design. For brevity, the user index $k$ is omitted in this section.
	
	\subsection{Problem Formulation}
	
	We aim to minimize the total transmit power at the AP by jointly optimizing
	the transmit beamforming at the AP and reflect beamforming at the IRS for  $K=1$. In this case, no inter-user interference is present in \eqref{r_se}, and thus the 
	problem is formulated as 
	\begin{align} 
	\mathrm{(P0)}:  
	\mathop{\mathtt{min}}_{\mathbf{w},\mathbf{v}, \{\theta_n\}}~&  \|\mathbf w\|^2 \label{eq:P0_Obj} \\
	\mathtt{s.t.}~&  |(\mathbf v^H  \mathbf \Phi + \mathbf h_d^H)\mathbf{w}|^2 \geq \gamma \sigma^2, \label{eq:P0_C1} \\
	~& v_n = \beta_n(\theta_n) e^{\jmath \theta_n},  n = 1,\dots,N, \label{eq:P0_C2} \\
	~& -\pi \leq \theta_n \leq \pi,  n = 1,\dots,N, \label{eq:P0_C3}
	\end{align}
	where $\gamma>0$ is the minimum SNR requirement of the user. Although the objective function of (P0) and constraints in \eqref{eq:P0_C3} are convex, it is challenging to solve (P0) due to the non-convex constraints in \eqref{eq:P0_C1} and \eqref{eq:P0_C2}. 	For any given $\mathbf{v}$ and $\{\theta_n\}_{n=1}^N$, it is not difficult to verify that the maximum-ratio transmission (MRT) is the optimal transmit beamforming solution to (P0), i.e., $\mathbf{w}^*=\sqrt{P}\frac{(\mathbf v^H  \mathbf \Phi + \mathbf h_d^H)^H}{\|\mathbf v^H  \mathbf \Phi + \mathbf h_d^H\|}$ \cite{wu2018intelligent}, where $P$ denotes the transmit power of the AP. By substituting $\mathbf{w}^*$ to (P0), the problem can be transformed to
	\begin{align}  
	\mathop{\mathtt{min}}_{P,\mathbf{v}, \{\theta_n\}}~& P \label{eq:P01_Obj} \\
	\mathtt{s.t.}~&  P\|\mathbf v^H  \mathbf \Phi + \mathbf h_d^H\|^2 \geq \gamma \sigma^2, \label{eq:P01_C1} \\
	~& v_n = \beta_n(\theta_n) e^{\jmath \theta_n},  n = 1,\dots,N, \label{eq:P01_C2} \\
	~& -\pi \leq \theta_n \leq \pi,  n = 1,\dots,N, \label{eq:P01_C3}
	\end{align}
	where  the optimal transmit power is given by $P^\star = \frac{\gamma \sigma^2}{\|\mathbf v^H  \mathbf \Phi + \mathbf h_d^H\|^2}$. Accordingly, the problem for minimizing the total transmit power at the AP by optimizing 
	$\mathbf{v}$ and $\{\theta_n\}_{n=1}^N$ can be equivalently  formulated as 
	\begin{align} 
	\mathrm{(P1)}:  
	\mathop{\mathtt{max}}_{\mathbf{v},\{\theta_n\}}~& \|\mathbf v^H  \mathbf \Phi + \mathbf h_d^H\|^2  \label{eq:P1_Obj} \\
	\mathtt{s.t.}~&  v_n = \beta_n(\theta_n) e^{\jmath \theta_n},  n = 1,\dots,N, \label{eq:P1_C1} \\
	~& -\pi \leq \theta_n \leq \pi,  n = 1,\dots,N. \label{eq:P1_C2}
	\end{align}
	It is worth noting that (P1) essentially corresponds to the effective channel power gain maximization  for the single user. Although simplified, problem (P1) is still non-convex and difficult to be optimally solved  due to the non-convex constraints in  \eqref{eq:P1_C1}. 

	\subsection{Power Loss of Ideal IRS Model with Asymptotically Large N}

	Prior to solving (P1), in this subsection, we first characterize the  receive power loss in a practical IRS-aided system when the phase shifts are designed by assuming the ideal  phase shift model, i.e., by setting $\beta_n(\theta_n)=1,\forall n$ in \eqref{eq:P1_C1}, as the number of IRS elements  $N\rightarrow \infty$. Since the IRS-reflected signal power dominates in the total received power for asymptotically large $N$, the signal received  from the AP-user link is ignored. To draw essential insight, we assume a single-antenna transmitter at the AP, i.e., $M=1$,  with the transmit power denoted by $P$.  Denote by $P_{\mathrm{ideal}}$ and $P_{\mathrm{practical}}$ the receive  power under the ideal and practical IRS phase shift models,  respectively, with the phase shifts obtained by solving (P1) by assuming the ideal IRS model, i.e., $\beta_n(\theta_n)=1,\forall n$ in \eqref{eq:P1_C1}.
	\begin{proposition} \label{qm_pro}
		Assume that $\mathbf h_{r}(n) \sim \mathcal{CN}(0,\,\varrho^{2}_r)$ and $\mathbf G(n,1) \sim \mathcal{CN}(0,\,\varrho^{2}_g)$, $ n = {1,\dots,N}$, and they are statistically independent.  As $N\rightarrow \infty$, we have
		\begin{equation} 
		\eta(\beta_{\text {min}},\alpha)=\frac{P_{\mathrm{practical}}}{P_{\mathrm{ideal}}} = \Bigg( \frac{1}{2\pi}\int_{-\pi}^{\pi} \beta(\theta) \ \ d\theta  \Bigg)^2, \label{loss}
		\end{equation}
		where $\beta(\theta)=(1-\beta_{\text {min}}) \left( \frac{\sin(\theta - \phi) +1}{2} \right)^\alpha + \beta_{\text {min}}$.
	\end{proposition} 
	\begin{IEEEproof}
		See Appendix A.
	\end{IEEEproof}

	From \eqref{loss}, it  is observed that as $N\rightarrow \infty$, the power ratio $\eta(\beta_{\text {min}},\alpha)$ depends only on $\beta_{\text {min}}$ and $\alpha$, but is regardless of $N$, which implies that the promising squared power scaling order, i.e., $\mathcal{O}(N^2)$ unveiled in \cite{wu2018intelligent} under the ideal phase shift model, still holds for practical phase shifters.   Besides,  it is observed from the numerical values in  Table I that $\eta(\beta_{\text {min}},\alpha)$ is more  sensitive to $\beta_{\text {min}}$ as compared to  $\alpha$. 
	For example, when $\beta_{\text {min}}=0.2$ and $\alpha=1.6$ (which  correspond to the  setup  in \cite{zhu2013active}),  a substantial power loss of $5.5$ dB is incurred under  the ideal IRS assumption. Even  for a much larger value of $\beta_{\text {min}}$, the power loss is still non-negligible, e.g.,  $\eta(0.8,1.6)=-1.1$ dB.
	These results suggest   that the consideration of  IRS hardware imperfection is indeed  crucial for the  beamforming design and achievable  performance   in practical systems. To this end, we solve (P1) in the next two subsections under the proposed practical phase shift model   by applying two different optimization   techniques, respectively.
	\begin{table}[!t] 
		\centering
		\caption {The Power Loss  Under Ideal IRS Assumption.}
		\begin{tabular}{|c|c|c|c|c|}
			\cline{1-5}
			$\eta(\beta_{\text {min}},\alpha)$ & $\beta_{\text {min}}=1.0$ & $\beta_{\text {min}}=0.8$ & $\beta_{\text {min}}=0.5$ & $\beta_{\text {min}}=0.2$ \\ \hline
			$\alpha=1.6$ & $0$ dB   & $-1.1$ dB   & $-3.0$ dB & $-5.5$ dB \\ \hline
			$\alpha=2.0$ & $0$ dB   & $-1.2$ dB  & $-3.2$ dB & $-6.0$ dB \\ \hline
		\end{tabular} \vspace{-0.8cm}
	\end{table}

	\subsection{AO-based Algorithm}
		
	First, we  propose an AO-based  algorithm to find an approximate solution to (P1), by iteratively optimizing  one phase shift  of the $N$ reflecting elements with those of the others  being fixed at each time,  until the objective value in \eqref{eq:P1_Obj} converges. 
	To this end, the problem for optimizing the reflection of the $n$-th element is simplified to
	\begin{align} 
	\mathrm{(P2)}:  
	\mathop{\mathtt{max}}_{\theta_n}~& \beta_n^2(\theta_n) \mathbf{\Psi}_{n,n} + \beta_n(\theta_n)|\varphi_n| \cos( \arg(\varphi_n)-\theta_n )  \label{eq:P1_AO_Obj}  \\
	\mathtt{s.t.}~&  -\pi \leq \theta_n \leq \pi, 
	\end{align}
	where $\mathbf{\Psi}=\diag(\mathbf h_r^H)\mathbf G\mathbf G^H\diag(\mathbf h_r)$, $\mathbf {\hat h}_d =\diag(\mathbf h_r^H)\mathbf G \mathbf h_d$, and $\varphi_n = \left( \sum_{m\neq n}^{N} \mathbf{\Psi}_{n,m}  v_m \right) +  2\hat{h}_{d,n} $. Note that \eqref{eq:P1_AO_Obj} is obtained by taking the terms associated with  $\beta_n(\theta_n)$ and $\theta_n$ in the expansion of \eqref{eq:P1_Obj}. It is noted that  (P2) is a single-variable non-convex optimization problem.
	
	Next, we propose a closed-form approximate solution to (P2). The key to approximately solving   (P2) in closed-form lies in re-expressing \eqref{eq:P1_AO_Obj} in a more tractable form.  In general, any approximation  of a  nonlinear function can only fit the original function values locally,  which we refer to as the trust region. For (P2), the trust region should  enclose its optimal solution, denoted by $\theta_n^*$.
	
	Define $f(\theta_n) \defeq \beta_n^2(\theta_n) \mathbf{\Psi}_{n,n} + \beta_n(\theta_n)|\varphi_n| \cos( \arg(\varphi_n)-\theta_n )$. It is not difficult to observe that for the ideal phase shift model considered in \cite{wu2018intelligent,qq_magazine,chongwang},  $\beta_n(\theta_n)$ and $\theta_n$  can be  designed to maximize $f(\theta_n) $ (or \eqref{eq:P1_AO_Obj})  by setting $\beta_n^*(\theta_n) =1$ and $\theta_n^*=\arg(\varphi_n), \forall n$. However, such a reflection design is no longer optimal  for a  practical IRS due to  the dependency of $\beta_n(\theta_n)$ on $\theta_n$ as depicted in Fig. \ref{reflection} (b). For instance, if $\arg(\varphi_n) = 0$, $\theta_n^* = 0$ may not be a favourable phase design as it yields  the lowest  reflection amplitude. In this case, $\theta_n^*$ needs to be properly chosen to balance  between $\beta_n(\theta_n) $ and  $\arg(\varphi_n)$. In particular, since the minimum $\beta_n(\theta_n)$ occurs  near zero  phase shift and  approaches the maximum  at  $\pi$ and $-\pi$, $\theta_n^*$ should  deviate from $\arg(\varphi_n)$ towards  $\pi$ (or $-\pi$) when $\arg(\varphi_n)$ is positive  (negative).  Based on this, the trust region of the optimal phase shift is presented in  the following proposition.
	\begin{proposition} \label{tr_pro}
		The trust region that encloses $\theta_n^*$ for (P2) is  given by 
		\begin{align} 
		\theta_n^* \in [\arg(\varphi_n),(-1)^\lambda\pi], \label{t_r}
		\end{align}
		with $\lambda =0$ when $\arg(\varphi_n) \geq 0$  and $\lambda =1$ otherwise.
	\end{proposition} 
	\begin{IEEEproof}
		See Appendix B.
	\end{IEEEproof}

	\begin{algorithm} [t!] 
		\caption{AO-based Algorithm  for Solving (P1)}\label{Tabel_2} 
		\begin{algorithmic}[1]
			\STATE \textbf{Initialize:} {  $\{\theta_n\}_{n=1}^N$. }
			\REPEAT
			\FOR{$n=1$ \textbf{to} $N$}
			\STATE Find $\theta_n^*$ as the solution to (P2) using \eqref{our_method} with  the trust region  given by \eqref{t_r}.
			\ENDFOR
			\STATE Obtain $v_n = \beta_n(\theta_n^*) e^{\jmath \theta_n^*},\forall n$.
			\UNTIL{{the objective value of (P1) with the obtained  $\mathbf{v}$ reaches convergence.}}
		\end{algorithmic}
	\end{algorithm}

	Motivated by the above result, a high-quality approximate solution to problem (P2) can be obtained numerically via the one-dimensional (1D) search over $[\arg(\varphi_n),(-1)^\lambda\pi]$, which may still be  computationally inefficient. Alternatively, a closed-form approximate solution  can be obtained by fitting a quadratic function  through three points over the trust region (which are obtained via equally sampling the trust  region), i.e., $\theta_A=\arg(\varphi_n)$, $\theta_B=\frac{\arg(\varphi_n)+(-1)^\lambda\pi}{2}$, and $\theta_C=(-1)^\lambda\pi$, as given in the following  proposition. 
	\begin{proposition} \label{sub_allocation_pro}
		Let	$f_1 = f(\theta_A)$, $f_2 = f(\theta_B)$, and	$f_3 = f(\theta_C)$. The approximate solution to (P2) obtained by fitting a quadratic function  through the points $(\theta_A,f_1)$, $(\theta_B,f_2)$, and $(\theta_C,f_3)$ is given by
		\begin{align} \label{our_method}
		\hat\theta_n^* = \frac{\theta_A(f_1-4f_2+3f_3)+\theta_C(3f_1-4f_2+f_3)  }{4(f_1-2f_2+f_3)}.
		\end{align}
	\end{proposition} 
	\begin{IEEEproof}
		See Appendix C.
	\end{IEEEproof}
	It is worth pointing out  that Proposition \ref{sub_allocation_pro} essentially corresponds to a single iteration of the  successive quadratic estimation with trust region refinement proposed in  \cite{powell}. 
	The overall AO-based  algorithm to solve (P1) is given in Algorithm 1. 

	\subsection{ Penalty-based Algorithm}
	
	To deal with the non-convex constraints in  \eqref{eq:P1_C1}, we next  resort to a penalty-based  method that  penalizes the constraint violation by adding a constraint-related penalty term to the objective function of (P1). To this end, the penalized version of (P1)  is formulated as
	\begin{align} 
	\mathrm{(P3)}:  
	\mathop{\mathtt{max}}_{\mathbf{v},\{\theta_n\}} ~& \|\mathbf v^H  \mathbf \Phi + \mathbf h_d^H\|^2 - \mu \sum_{n=1}^{N} |v_n-\beta_n(\theta_n) e^{\jmath \theta_n}|^2 \label{eq:P2_Obj} \\
	~& -\pi \leq \theta_n \leq \pi,  n = 1,\dots,N, \label{eq:P2_C2}
	\end{align}
	where $\mu>0$ is the penalty parameter that imposes a  cost for the constraint violation of the constraints in \eqref{eq:P1_C1}. In particular, when $\mu \rightarrow \infty$, solving the above problem yields an approximate solution to  (P1) \cite{penalty_method}. 	However, initializing $\mu$  to be a sufficiently small value generally yields a good starting point for the proposed algorithm, even though  this point may be infeasible for (P1). By gradually increasing the value of $\mu$ by a factor of $\varrho>1$, we can maximize the original objective function, i.e., $\|\mathbf v^H  \mathbf \Phi + \mathbf h_d^H\|^2$, and  obtain a solution that satisfies all the equality constraints in  \eqref{eq:P1_C1} within a predefined accuracy. This thus leads to  a two-layer iterative algorithm, where the inner layer solves the penalized optimization problem (P3) while the outer layer updates the penalty coefficient $\mu$, until the convergence is achieved.  
	
	For any given $\mu>0$, (P3) is still a non-convex optimization problem due to the non-convex objective function. However, it is observed from \eqref{eq:P2_Obj} that  $\mathbf v$  can be updated  with  fixed $\{\theta_n\}_{n=1}^N$, and  $\{\theta_n\}_{n=1}^N$ can be updated in parallel with fixed $\mathbf v$, which thus motivates us to apply the block coordinate descent (BCD) method to solve (P3) efficiently. The convergence is achieved when the fractional increase of \eqref{eq:P2_Obj} is below a positive yet sufficiently small threshold and the details are given as follows.  
	
	1) For any given $\{\theta_n\}_{n=1}^N$, $\mathbf v$ in (P3) can be optimized by solving the following problem
	\begin{align} 
	\mathrm{(P3.1)}:  
	\mathop{\mathtt{max}}_{\mathbf{v}} ~& \|\mathbf v^H  \mathbf \Phi + \mathbf h_d^H\|^2 - \mu \|\mathbf{v} - \mathbf{a} \|^2, \label{eq:P2.1_Obj} 
	\end{align}
	where $ \mathbf {a} = [\beta_1(\theta_1) e^{\jmath \theta_1}, \dots, \beta_N(\theta_N) e^{\jmath \theta_N}]^T$. It is not difficult to observe that (P3.1) is an unconstrained non-convex optimization problem due to the first and  second terms of \eqref{eq:P2.1_Obj} being respectively convex and concave, for which  we can apply the concave-convex procedure (CCCP)  \cite{mm_tsp} to approximately solve it in an iterative manner. Specifically, at each iteration $l=1,2,\dots,$ we approximate the first term of the objective function in (P3.1) by a linear function using its first-order Taylor series at a given point $ \mathbf v^{(l)} $ to form a convex approximate optimization problem, which is given by (with constant terms ignored)
	\begin{align} 
	\mathrm{(P3.2)}:  
	\mathop{\mathtt{max}}_{\mathbf{v}} ~& 2(\mathbf \Phi \mathbf \Phi^H \mathbf v^{(l)} + \mathbf \Phi \mathbf h_d)^H(\mathbf v-\mathbf v^{(l)}) - \mu \|\mathbf{v} - \mathbf{a} \|^2.  \label{eq:P2.1_Obj2} 
	\end{align}
	 Next, we set the value of  $ \mathbf v $ for iteration $l+1$ as the optimal solution to (P3.2) at iteration $l$, and the algorithm continues until the objective value of (P3.1) reaches convergence. It is not difficult to observe that (P3.2) is an unconstrained  convex optimization problem, for which the optimal solution in closed-form can be easily obtained as (by setting the first-order derivative of the objective function with respect to $\mathbf v$ equal to zero)
	\begin{align} \label{update_v}
	\mathbf{v}^{(l+1)} &= \frac{\mathbf \Phi \mathbf \Phi^H \mathbf  v^{(l)} + \mathbf \Phi \mathbf h_d + \mu \mathbf {a}}{\mu}.
	\end{align}

	\vspace{5mm}
	
	2) For any given $\mathbf v$, $\{\theta_n\}_{n=1}^N$ in (P3) can be optimized by solving the following problem	
	\begin{align}   
	\mathrm{(P3.3)}:
	\mathop{\mathtt{max}}_{\{\theta_n\}}~& -\sum_{n=1}^{N}|v_n-\beta_n(\theta_n) e^{\jmath \theta_n}|^2 \label{update_teta11}\\
	~& -\pi \leq \theta_n \leq \pi,  n = 1,\dots,N. 
	\end{align} 
	It is noted that this is a non-convex optimization problem due to the fact that  $\beta_n(\theta_n) $ is non-convex in $\theta_n$. However, since $\theta_n$'s are fully separable in the objective function,  solutions to $\{\theta_n\}_{n=1}^N$ can be obtained  by solving $N$ independent subproblems in parallel. By expanding $|v_n-\beta_n(\theta_n) e^{\jmath \theta_n}|^2$ and ignoring constant terms, each  corresponding subproblem is given by	
	\begin{align} 
	\mathrm{{(P3.4)}}: 
	\mathop{\mathtt{max} }_{\theta_n} ~& 2\beta_n(\theta_n)|v_n|\cos(\psi_n-\theta_n) - \beta_n^2(\theta_n) \label{update_teta22}\\
	~& -\pi \leq \theta_n \leq \pi. 
	\end{align}
	where $\psi_n=\arg(v_n)$.
	
		\begin{algorithm} [t!]
		\caption{Penalty-based Algorithm for Solving (P1)}\label{Tabel_1}
		\begin{algorithmic}[1]
			\STATE \textbf{Initialize:} {  $\{\theta_n\}_{n=1}^N$ and $\mu>0$. }
			\REPEAT
			\REPEAT
			\STATE Update $\mathbf{v}$ as the solution to (P3.1).
			\STATE Update $\{\theta_n\}_{n=1}^N$ using \eqref{our_method} with  the trust region  given by \eqref{pro3eq}.
			\UNTIL{{The fractional increase  of the objective value of (P3) is below a threshold $  \epsilon_1>0$ or the maximum number of iterations is reached.}	}
			\STATE Update the penalty coefficient $\mu \leftarrow \varrho\mu$.
			\UNTIL{{The constraint violation ($\sum_{n=1}^{N} |v_n-\beta_n(\theta_n) e^{\jmath \theta_n}|^2$)  is below a threshold $  \epsilon_2>0$.}}
		\end{algorithmic}
	\end{algorithm}
	
	In contrast to (P2), the trust region given in Proposition \ref{tr_pro} is not applicable for  (P3.4) due to the negative term in its objective function, i.e., $- \beta_n^2(\theta_n)$. Nevertheless, it can be observed from \eqref{update_teta22} that its $\cos(\cdot)$ term is maximized when $\theta_n = \psi_n$ and the whole function is maximized when $\theta_n$ slightly deviates away from $\psi_n$ based on   $v_n$. The trust region is thus formally presented in the following proposition.
	\begin{proposition} \label{tr_pro2}
		The trust region that encloses the optimal solution of (P3.4), denoted by $\theta_n^*$, is  given by 
		\begin{align}
		\theta_n^* \in 
		\begin{cases} 
		[\psi_n,\psi_n+(-1)^\lambda\Delta] & \text{if } \frac{\beta_n(\psi_n)+\beta_n(\psi_n+\Delta)}{2} < |v_n|, \\
		[\psi_n,\psi_n-(-1)^\lambda\Delta]       & \text{if } \frac{\beta_n(\psi_n)+\beta_n(\psi_n-\Delta)}{2} > |v_n|,
		\end{cases} \label{pro3eq}
		\end{align}
		where $\Delta \geq 0$,  and $\lambda =0$ when $\psi_n \geq 0$  and $\lambda =1$ otherwise.
	\end{proposition} 
	\begin{IEEEproof}
		See Appendix D.
	\end{IEEEproof}
	By choosing a proper value for $\Delta$ and fitting a quadratic function  through three points over the trust region (which are obtained via equally sampling the trust  region), a closed-form approximate solution to (P3.4) can be similarly obtained as Proposition \ref{sub_allocation_pro}. The overall penalty-based algorithm to solve (P1) is given in Algorithm 2.

	\section{Multiuser Beamforming Optimization} \label{mu_sec}
	
	In this section, we consider the general multiuser setup. Specifically, we propose two efficient algorithms to solve the multiuser beamforming optimization problem suboptimally, by extending the solutions for the  single-user case.
	
	\subsection{Problem Formulation}
	
	We aim to minimize the total transmit power at the AP by jointly optimizing 	the transmit beamforming at the AP and reflect beamforming at the IRS, subject to the  individual SINR constraints at all users. Accordingly, the problem is formulated as	
	\begin{align} 
	\mathrm{(P4)}:  
	\mathop{\mathtt{min}}_{ \{\mathbf{w}_k\},\mathbf{v}, \{\theta_n\}}~& \sum_{k=1}^{K} \|\mathbf{w}_k\|^2 \label{eq:P4_Obj} \\ 
	\mathtt{s.t.}  &   \frac{|(\mathbf v^H  \mathbf \Phi_k + \mathbf h_{d,k}^H)\mathbf{w}_k|^2}{\sum_{j \neq k}^{K} |(\mathbf v^H  \mathbf \Phi_k + \mathbf h_{d,k}^H)\mathbf{w}_j|^2 + \sigma_k^2} \geq \gamma_k,  k=1,\dots,K, \label{eq:P4_C1} \\
	 &  v_n = \beta_n(\theta_n) e^{\jmath \theta_n},  n = 1,\dots,N, \label{eq:P4_C2} \\
	 &   -\pi \leq \theta_n \leq \pi,  n = 1,\dots,N, \label{eq:P4_C3}
	\end{align}
	where $\gamma_k>0$ is the minimum SINR requirement of user $k$. Note that  (P4) is a non-convex optimization problem due to the coupling between $\mathbf{w}_k$'s and $\mathbf v$ in \eqref{eq:P4_C1} and non-convex constraints in \eqref{eq:P4_C2}, for which we propose two efficient algorithms by generalizing the two approaches in the single-user case.
		
	\subsection{Extended Penalty-based Algorithm} \label{epb}
	
	First, we introduce new auxiliary variables to decouple $\mathbf{w}_k$'s and $\mathbf v$ in \eqref{eq:P4_C1}. To this end, let $\mathbf h_k^H\mathbf{w}_j = x_{k,j}$ with $\mathbf h_k^H = \mathbf v^H  \mathbf \Phi_k + \mathbf h_{d,k}^H, k,j=1,\dots,K$. Then the SINR constraints can be expressed as
	\begin{align} \label{new_sinr}
	\frac{|x_{k,k}|^2}{\sum_{j \neq k}^{K} |x_{k,j}|^2 + \sigma_k^2} \geq \gamma_k,  k=1,\dots,K. 
	\end{align}
	By replacing \eqref{eq:P4_C1} with \eqref{new_sinr}, (P4) is equivalently transformed to
	\begin{align} 
	\mathrm{(P5)}:  
	\mathop{\mathtt{min}}_{ \{\mathbf{w}_k\},\mathbf{v}, \{\theta_n\},\{x_{k,j}\}}~& \sum_{k=1}^{K} \|\mathbf{w}_k\|^2 \label{eq:P5_Obj} \\ 
	\mathtt{s.t.}  &   \frac{|x_{k,k}|^2}{\sum_{j \neq k}^{K} |x_{k,j}|^2 + \sigma_k^2} \geq \gamma_k,  k=1,\dots,K, \label{eq:P5_C1} \\
	&  v_n = \beta_n(\theta_n) e^{\jmath \theta_n},  n = 1,\dots,N, \label{eq:P5_C2} \\
	& \mathbf h_k^H\mathbf{w}_j = x_{k,j},  k,j=1,\dots,K,  \label{eq:P5_C3} \\
	&   -\pi \leq \theta_n \leq \pi,  n = 1,\dots,N. \label{eq:P5_C4}
	\end{align}	
	(P5) is still non-convex and difficult to be optimally solved in general due to the non-convex constraints in \eqref{eq:P5_C2} and  the coupling between  $\mathbf{w}_k$'s and $\mathbf v$  is still present with  the newly added equality constraint in \eqref{eq:P5_C3}. To overcome such difficulty, we resort to the penalty-based method  by adding equality constraint-related penalty terms to the objective function of (P5), yielding the following optimization problem
	\begin{align} 
	\mathrm{(P6)}:  
	\mathop{\mathtt{min}}_{ \{\mathbf{w}_k\},\mathbf{v}, \{\theta_n\},\{x_{k,j}\}}~& \sum_{k=1}^{K} \|\mathbf{w}_k\|^2 + \mu \sum_{n=1}^{N} |v_n-\beta_n(\theta_n) e^{\jmath \theta_n}|^2 + \nu \sum_{k=1}^{K} \sum_{j=1}^{K} |\mathbf h_k^H\mathbf{w}_j-x_{k,j}|^2 \label{eq:P6_Obj} \\ 
	\mathtt{s.t.}  &   \frac{|x_{k,k}|^2}{\sum_{j \neq k}^{K} |x_{k,j}|^2 + \sigma_k^2} \geq \gamma_k,  k=1,\dots,K, \label{eq:6_C1} \\
	 &   -\pi \leq \theta_n \leq \pi,  n = 1,\dots,N, \label{eq:P6_C3}
	\end{align}
	where $\mu>0$ and $\nu>0$ denote the penalty coefficients used for penalizing the violation of equality constraints in (P5). Similarly to  problem (P3), we propose a two-layer iterative algorithm. Specifically, the inner layer solves the penalized optimization problem (P6) by applying the BCD method while the outer layer updates $\mu$ and $\nu$, until the convergence is achieved. 
	To this end, the entire optimization variables in (P6) are partitioned into four blocks, i.e.,  $\{\mathbf{w}_k \}_{k=1}^K$,  $\mathbf v$,  $\{\theta_n\}_{n=1}^N$, and $\{x_{k,j}\}_{k,j=1}^K$. Then, we can minimize  (P6) by alternately optimizing each of the above four blocks in one iteration with the other three blocks fixed, and iterating the above until the convergence is reached, which is detailed as follows. \vspace{5mm}
	
	1) For any given $\mathbf v$, $\{\theta_n\}_{n=1}^N$ and $\{x_{k,j}\}_{k,j=1}^K$, $\{\mathbf{w}_k \}_{k=1}^K$ can be optimized by solving the following problem
	\begin{align} 
	\mathrm{(P6.1)}:  
	\mathop{\mathtt{min}}_{ \{\mathbf{w}_k\}}~& \sum_{k=1}^{K} \|\mathbf{w}_k\|^2 + \nu \sum_{k=1}^{K} \sum_{j=1}^{K} |\mathbf h_k^H\mathbf{w}_j-x_{k,j}|^2. \label{eq:P6.1_Obj} 
	\end{align}
	It is not difficult to observe that (P6.1) is an unconstrained  convex optimization  problem, for which the optimal solution in closed-form can be easily obtained by setting the first-order derivative of the objective function with respect to $\mathbf w_k$ equal to zero, and is given by
	\begin{align} \label{mu_update_w}
	\mathbf{w}_k^* &=  \nu \Big( \mathbf I_M + \nu \sum_{k=1}^{K} \mathbf h_k \mathbf h_k^H \Big)^{-1} \Big( \sum_{j=1}^{K} \mathbf h_j x_{j,k} \Big).
	\end{align}
	It is worth pointing out that all $\mathbf{w}_k^*$'s for different users can be updated in parallel by using \eqref{mu_update_w}.
	
	\vspace{5mm}
	
	2) For any given $\{\mathbf{w}_k \}_{k=1}^K$,  $\{\theta_n\}_{n=1}^N$ and $\{x_{k,j}\}_{k,j=1}^K$, $\mathbf v$ can be optimized by solving the following problem
	\begin{align} 
	\mathrm{(P6.2)}:  
	\mathop{\mathtt{min}}_{ \mathbf{v}}~&  \mu  \|\mathbf v-\mathbf a \|^2 + \nu \sum_{k=1}^{K} \sum_{j=1}^{K} |\mathbf h_k^H\mathbf{w}_j-x_{k,j}|^2, \label{eq:P6.2_Obj}
	\end{align}
	where $ \mathbf {a} = [\beta_1(\theta_1) e^{\jmath \theta_1}, \dots, \beta_N(\theta_N) e^{\jmath \theta_N}]^T$. By applying the change of variables, $\diag(\mathbf h_{r,k}^H) \mathbf G \mathbf{w}_j= \mathbf{\bar d}_{k,j} $ and $-\mathbf h_{d,k}^H\mathbf{w}_j + x_{k,j} = \bar C_{k,j}$, we have
	\begin{align} 
	\mathbf h_k^H\mathbf{w}_j-x_{k,j} = \mathbf v^H  \diag(\mathbf h_{r,k}^H) \mathbf G \mathbf{w}_j + \mathbf h_{d,k}^H\mathbf{w}_j - x_{k,j}  = \mathbf v^H\mathbf{\bar d}_{k,j}-\bar C_{k,j}.
	\end{align}
	Problem (P6.2) is thus equivalent to 
	\begin{align} 
	\mathrm{(P6.2-EQ)}:  
	\mathop{\mathtt{min}}_{ \mathbf{v}}~&  \mu  \|\mathbf v-\mathbf a \|^2 + \nu \sum_{k=1}^{K} \sum_{j=1}^{K} |\mathbf v^H\mathbf{\bar d}_{k,j}-\bar C_{k,j}|^2, \label{eq:P6.2eq_Obj}
	\end{align}
	for which the optimal solution in closed-form can be  similarly obtained  by setting the first-order derivative of the objective function with respect to $\mathbf v$ equal to zero, and is given by
	\begin{align} \label{mu_update_v}
	\mathbf{v}^* &=  \Big( \mu\mathbf I_N + \nu \sum_{k=1}^{K}\sum_{j=1}^{K}\mathbf{\bar d}_{k,j}\mathbf{\bar d}_{k,j}^H \Big)^{-1} \Big( \mu\mathbf a + \nu \sum_{k=1}^{K}\sum_{j=1}^{K}\mathbf{\bar d}_{k,j}\bar C_{k,j}^H \Big).
	\end{align}
	
	\vspace{5mm}

	3) For any given $\{\mathbf{w}_k \}_{k=1}^K$, $\mathbf v$ and $\{x_{k,j}\}_{k,j=1}^K$, $\{\theta_n\}_{n=1}^N$  can be optimized by solving the following problem
	\begin{align} \label{update_teta1}
	\mathrm{(P6.3)}:  
	\mathop{\mathtt{min}}_{\{\theta_n\}}~& \sum_{n=1}^{N}|v_n-\beta_n(\theta_n) e^{\jmath \theta_n}|^2,
	\end{align} 
	which is essentially identical to (P3.3) and thus can be similarly solved (i.e., by parallelly solving $N$ independent subproblems in closed-form) according to  Proposition \ref{sub_allocation_pro}, while the trust region is given by Proposition \ref{tr_pro2}. 
	
	\vspace{5mm}
	
	4) For any given $\{\mathbf{w}_k \}_{k=1}^K$, $\mathbf v$ and $\{\theta_n\}_{n=1}^N$, $\{x_{k,j}\}_{k,j=1}^K$ can be optimized by solving the following problem
	\begin{align} 
	\mathrm{(P6.4)}:  
	\mathop{\mathtt{min}}_{ \{x_{k,j}\}}~&  \sum_{k=1}^{K} \sum_{j=1}^{K} |\mathbf h_k^H\mathbf{w}_j-x_{k,j}|^2 \label{eq:P6.4_Obj} \\ 
	\mathtt{s.t.}  &   \frac{|x_{k,k}|^2}{\sum_{j \neq k}^{K} |x_{k,j}|^2 + \sigma_k^2} \geq \gamma_k,  k=1,\dots,K. \label{eq:6.4_C1} 
	\end{align}
	It is not difficult to observe that the optimization
	variables with respect to different users are separable in both the objective function and constraints. As a result,  we can solve (P6.4) by solving $K$ independent subproblems in
	parallel, each with only one single SINR constraint. To this end,  the corresponding subproblem for the $k$-th user  with respect to $x_{k,j}$'s, $\forall j=1,\dots,K$, can be given by
	\begin{align} 
	\mathrm{{(P6.5)}}:  
	\mathop{\mathtt{min}}_{ \{x_{k,j},\forall j\}}~&   \sum_{j=1}^{K} |\mathbf h_k^H\mathbf{w}_j-x_{k,j}|^2 \label{eq:P6.4sub_Obj} \\ 
	\mathtt{s.t.}  &   \frac{|x_{k,k}|^2}{\sum_{j \neq k}^{K} |x_{k,j}|^2 + \sigma_k^2} \geq \gamma_k. \label{eq:6.4sub_C1} 
	\end{align}
	Although (P6.5) is  non-convex, it has been shown in \cite{IRS_SWIPT_2} that this problem can be efficiently and optimally solved  by applying the Lagrange duality method. Specifically,  by exploiting the first-order optimality condition,  the optimal solution is given by
	\begin{align} 
	x_{k,k}^* &= \frac{\mathbf h_k^H\mathbf{w}_k}{1-\lambda_k}, \label{dual1}\\
	x_{k,j}^* &= \frac{\mathbf h_k^H\mathbf{w}_j}{1+\lambda_k \gamma_k}, j\neq k,k=1,\dots,K, \label{dual2}
	\end{align}
	where $\lambda_k$ is the dual variable. If the SINR constraint in \eqref{eq:6.4sub_C1} is not met with equality at the optimal solution, then  $\lambda_k=0$. Otherwise, the optimal dual variable can be efficiently obtained by substituting \eqref{dual1}-\eqref{dual2}  into \eqref{eq:6.4sub_C1}, i.e., $\frac{|\mathbf h_k^H\mathbf{w}_k|^2}{(1-\lambda_k)^2} - \sum_{j\neq k}^{K} \frac{\gamma_k |\mathbf h_k^H\mathbf{w}_j|^2}{(1+\lambda_k \gamma_k)^2}-\gamma_k\sigma_k^2=0$, and performing a simple bisection search over $0\leq \lambda_k <1 $. 
	
	\begin{algorithm} [t!]
		\caption{Extended Penalty-based Algorithm for Solving (P4)}\label{Tabel_3}
		\begin{algorithmic}[1]
			\STATE \textbf{Initialize:} { $\mathbf{v}$, $\{\theta_n\}_{n=1}^N$, $\{x_{k,j}\}_{k,j=1}^K$, $\mu>0$, and $\nu>0$ }
			\REPEAT
			\REPEAT
			\STATE Update $\{\mathbf{w}_k\}_{k=1}^K$ by solving (P6.1).
			\STATE Update $\mathbf{v}$ by solving (P6.2).
			\STATE Update $\{\theta_n\}_{n=1}^N$ by solving (P6.3).
			\STATE Update $\{x_{k,j}\}_{k,j=1}^K$ by solving (P6.4).
			\UNTIL{{The fractional decrease of the objective value of (P6) is below a threshold $  \epsilon_1>0$ or the maximum number of iterations is reached.}	}
			\STATE Update the penalty coefficients as $\mu \leftarrow \varrho\mu$ and $\nu \leftarrow \varrho\nu$.
			\UNTIL{{The constraint violation   is below a threshold $  \epsilon_2>0$ or  $\sum_{k=1}^{K} \|\mathbf{w}_k\|^2$ reaches convergence.}}
		\end{algorithmic}
	\end{algorithm}
	
	The overall penalty-based algorithm to solve (P4) is given in Algorithm 3.  It should be noted that  Algorithm 3  is computationally efficient as the optimization variables in steps  4, 5, and 6 can be updated in parallel by using closed-form expressions, and those in step 6 can be obtained by using the simple bisection search. In particular, it can be shown that the complexity of solving (P6.1) is $ \mathcal{O}(K(N^2+NM+M^2)+M^3)$, that of solving (P6.2) is $ \mathcal{O}(K^2(N^2+NM+M^2)+N^3)$, that of solving (P6.3) is $ \mathcal{O}(1)$, and that of solving (P6.4) is $ \mathcal{O}(K^2\log_2(1/\epsilon_3))$ where $\epsilon_3$ denotes  the accuracy for the bisection search. Thus, the overall complexity of Algorithm 3 is given by $\mathcal{O} (  I_{{inn}} I_{{out}} (N^3+M^3+K^2(N^2+NM+M^2)+K^2\log_2(1/\epsilon_3))   )$ where $I_{{inn}}$ and $I_{{out}}$  respectively denote the inner and outer iteration numbers required for convergence.
	
	\subsection{Two-Stage Algorithm}
	
	Inspired by the combined channel power gain maximization problem (P1) for the single-user case, we next propose a two-stage algorithm with lower complexity as compared to the penalty-based algorithm. Specifically, the phase shifts at the IRS are optimized in the first stage by solving the following weighted effective channel power gain maximization problem.
	\begin{align} 
	\mathrm{(P7.1)}:  
	\mathop{\mathtt{max}}_{\mathbf{v},\{\theta_n\}}~& \sum_{k=1}^{K} \Psi_k \|(\mathbf v^H  \mathbf \Phi_k + \mathbf h_{d,k}^H)\|^2  \label{eq:P7_Obj} \\
	\mathtt{s.t.}~&  v_n = \beta_n(\theta_n) e^{\jmath \theta_n},  n = 1,\dots,N, \label{eq:P6_C1} \\
	~& -\pi \leq \theta_n \leq \pi,  n = 1,\dots,N, \label{eq:P7_C2}
	\end{align}
	where we set the weights to be $\Psi_k=\frac{1}{\gamma_k \sigma_k^2},k=1,\dots,K$, motivated by the constraint in \eqref{eq:P4_C1}. This aims to align  the phases of different user channels so as to maximize the active and passive beamforming gains of the system.  (P7.1) can be similarly solved by adopting the penalty-based technique proposed in Section \ref{su}, thus the details are  omitted  for brevity.
	
	In the second stage, we solve the following problem  to obtain the optimal transmit beamforming with the phase shifts obtained from (P7.1), 
	\begin{align} 
	\mathrm{(P7.2)}:  
	\mathop{\mathtt{min}}_{ \{\mathbf{w}_k\}}~& \sum_{k=1}^{K} \|\mathbf{w}_k\|^2  \label{eq:P5.1_Obj} \\ 
	\mathtt{s.t.} ~&  \frac{|\mathbf h_k^H\mathbf{w}_k|^2}{\sum_{j \neq k}^{K} |\mathbf h_k^H\mathbf{w}_j|^2 + \sigma_k^2} \geq \gamma_k,  k=1,\dots,K, \label{eq:P5.1_C1} 
	\end{align}
	where $\mathbf h_k^H = \mathbf v^H  \mathbf \Phi_k + \mathbf h_{d,k}^H $. Note that (P7.2) is  the conventional power minimization problem in the multiuser MISO downlink broadcast channel, where its optimal solution known as the minimum mean squared error (MMSE) based linear precoder can be obtained by using, e.g., a  fixed-point iteration algorithm based on the uplink-downlink duality \cite{wu2018intelligent,1262126,1664998}, i.e.,	
	\begin{align} 
	\mathbf{w}_k^* = \sqrt{p_k} \mathbf{\hat w}_k^*,  \label{mmse1}
	\end{align}
	where 
	\begin{align}
	\begin{bmatrix}
	p_1 \\
	\vdots \\
	p_K
	\end{bmatrix} = \mathbf{Q}^{-1}
	\begin{bmatrix}
	\sigma_1^2 \\
	\vdots \\
	\sigma_K^2 
	\end{bmatrix}, \label{mmse2}
	\end{align}
	\begin{align}
	\mathbf{Q}(i,j) = 
	\begin{cases} 
	\frac{1}{\gamma_i} |\mathbf h_i^H \mathbf{\hat w}_i^*|^2, &  i=j, \\
	-|\mathbf h_i^H \mathbf{\hat w}_j^*|^2,  &  i\neq j, \forall i,j \in \{1,\dots,K\}, \label{mmse3}
	\end{cases}
	\end{align}
	\begin{align} 
	\mathbf{\hat w}_k^* = \frac{(\mathbf I_M + \sum_{i=1}^{K}\frac{\rho_i}{\sigma_i^2}\mathbf h_i\mathbf h_i^H)^{-1}\mathbf h_k}{\|(\mathbf I_M + \sum_{i=1}^{K}\frac{\rho_i}{\sigma_i^2}\mathbf h_i\mathbf h_i^H)^{-1}\mathbf h_k\|},  \label{mmse4}
	\end{align}
	\begin{align} 
	\rho_k = \frac{\sigma_k^2}{(1+\frac{1}{\gamma_k}) \mathbf h_k^H (\mathbf I_M + \sum_{i=1}^{K}\frac{\rho_i}{\sigma_i^2}\mathbf h_i\mathbf h_i^H)^{-1}\mathbf h_k},  k=1,\dots,K. \label{mmse5}
	\end{align}
	Specifically, $\rho_k$'s can be obtained by using the fixed-point algorithm to solve $K$ equations in \eqref{mmse5}. With $\rho_k$'s, $\mathbf{\hat w}_k^*$'s can be obtained from \eqref{mmse4} and then $p_k$'s can be obtained from \eqref{mmse2}. Finally, $\mathbf{w}_k^*$'s are obtained by using \eqref{mmse1} with $\mathbf{\hat w}_k^*$'s and $p_k$'s.	
	
	The overall complexity of the two-stage algorithm can be shown to be $ \mathcal{O}( I_{itr}(KM^2+M^3)+K^3+K^2M+KMN + I_{{inn}} I_{{out}}(N^3+KN^2)) $ where $I_{itr}$ denotes the number of iterations required for obtaining  $\rho_k$'s in \eqref{mmse5}, and $I_{{inn}}$, $I_{{out}}$  respectively denote the inner and outer iteration numbers required for the convergence of (P7.1). Compared to the extended penalty-based algorithm proposed in Section \ref{epb}, the two-stage algorithm has lower computational complexity as (P7.1) and (P7.2) only need to be respectively solved for one time.
		
	\section{Simulation Results} \label{sr_sec}
	
	\begin{figure}[t!] 
		\centering \vspace{-0.5cm}
		\includegraphics[trim = 0mm 0mm 0mm 0mm, clip,width=0.5\linewidth]{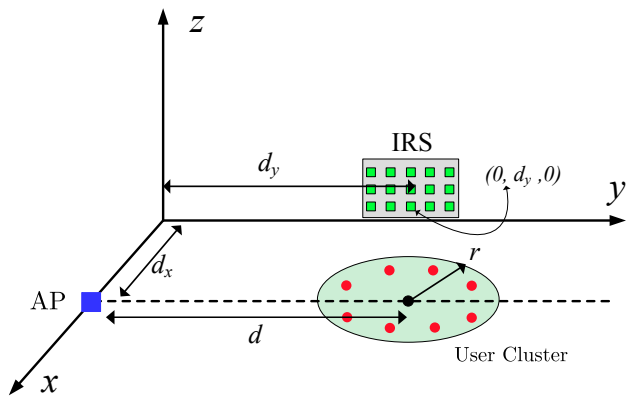}
		\caption{Simulation setup.}  
		\label{su_sm} \vspace{-0.8cm}
	\end{figure}

	In this section, numerical results  are provided to evaluate the performance of the proposed algorithms. We consider a system that operates on a carrier frequency of $2.4$ GHz, which corresponds to the signal attenuation at a reference distance of $1$ m about $40$ dB. A three-dimensional (3D) coordinate system  is considered as shown in Fig. \ref{su_sm}, where a uniform linear array (ULA) at the AP and a uniform rectangular array (URA) at the IRS are located in $x$-axis and $y$-$z$ plane, respectively. The reference antenna/element at the AP/IRS are respectively located at $(d_x,0,0)$ and $(0,d_y,0)$, where in both cases a half-wavelength (i.e., $6.25$ centimeter) spacing is assumed among adjacent antennas/elements. For the IRS, we set $N=N_y N_z$ where $N_y$ and $N_z$ denote the number of reflecting elements along $y$-axis and $z$-axis, respectively. For the 	purpose of exposition, we fix $N_y=5$ and increase $N_z$ linearly with $N$.
	The $K$ users are uniformly and randomly distributed in a cluster, which is centered at $(d_x,d,0)$ with radius $r$. Rayleigh fading is assumed for all the channels involved. The path loss exponents are set to $2.2$, $2.8$, and $3.8$ for the channels between AP-IRS, IRS-user, and AP-user, respectively, as IRSs are usually deployed for users with weak AP-user channels and their
	locations can be properly selected to avoid severe blockage with the AP. Moreover, we set $\sigma_k^2=-94$ dBm, $\forall k$. The  phase shift model parameters are set as follows unless specified otherwise: $\beta_{\text{min}}=0.2$, $\alpha=1.6$, and $\phi=0.43\pi$ according to  \cite{zhu2013active}.

	\subsection{Single-User Case} \label{simulation1}
	
	We first   consider a single-user system with the SNR target $\gamma = 10$ dB and $M=4$. The user is assumed to lie in  the cluster center, denoted by  $(d_x,d,0)$ with $d_x=2$ m. Moreover, it is assumed that $d_y=400$ m.


	\begin{figure}
		\centering \vspace{-0.8cm}
		\begin{minipage}{.48\textwidth} 
			\includegraphics[width=\textwidth,height=\textheight,keepaspectratio]{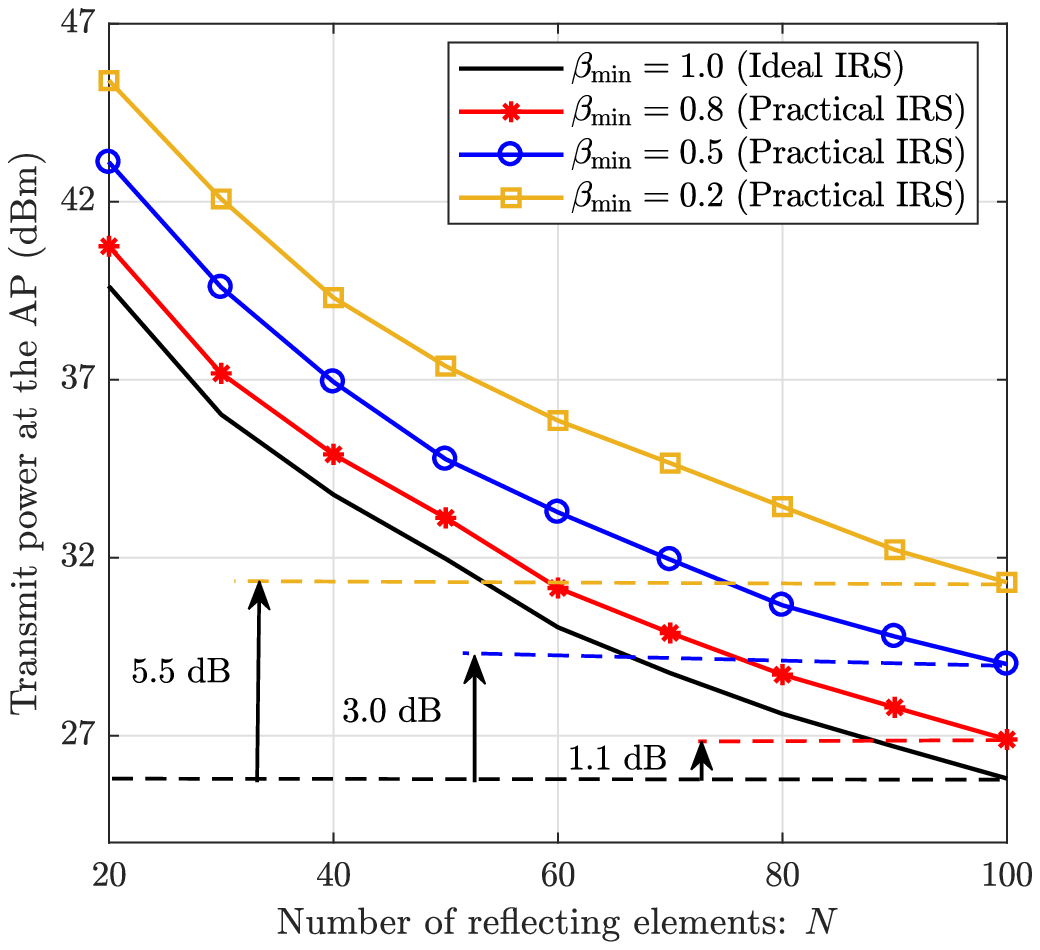}
			\caption{AP transmit power versus  number of reflecting elements.}   
			\label{prop_1_validate} 
		\end{minipage}
		\hspace{0.3cm}
		\begin{minipage}{.48\textwidth} 
			\includegraphics[width=\textwidth,height=\textheight,keepaspectratio]{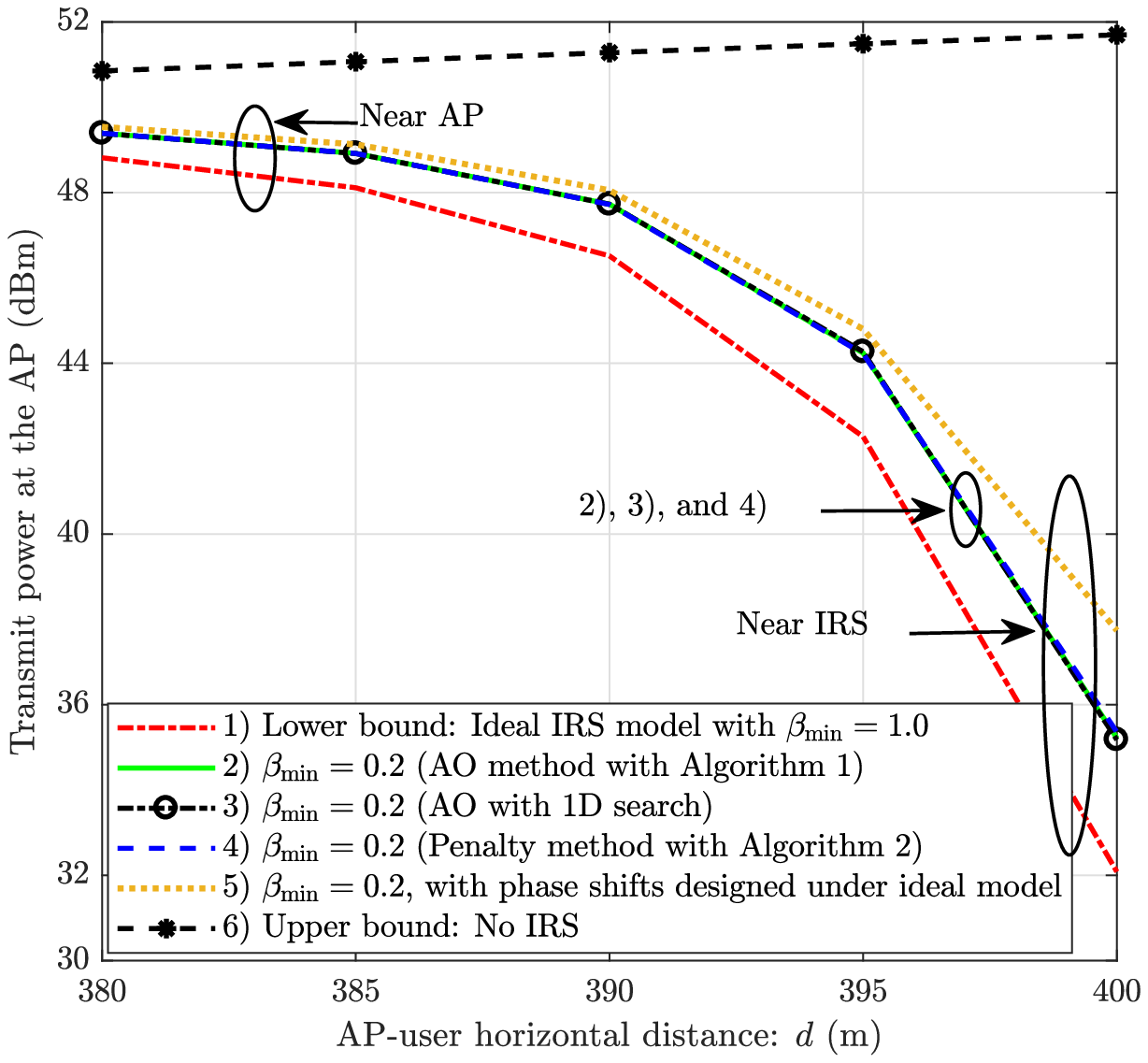}
			\caption{AP transmit power versus the AP-user horizontal distance.}
			\label{variable_d_power} 
		\end{minipage} \vspace{-0.8cm}
	\end{figure}
	
	To validate the theoretical analysis in Proposition \ref{qm_pro}, we plot in Fig. \ref{prop_1_validate} the AP transmit power  versus the number of reflecting elements $N$  when $d=400$ m. In particular, IRS phase shifts are obtained for the ideal IRS (i.e., $\beta_{\text{min}}=1$) by using Algorithm 1 and the obtained  phase shifts are then applied to the practical IRS with $\beta_{\text{min}}=0.8$, $\beta_{\text{min}}=0.5$, or $\beta_{\text{min}}=0.2$. Since the power loss is not sensitive to  parameter $\alpha$ in the phase shift model, we fix $\alpha=1.6$ in simulations. From Fig. \ref{prop_1_validate}, it is observed that as $N$ increases, the performance gap between the ideal case ($\beta_{\text{min}}=1$) and practical cases ($\beta_{\text{min}}=0.8$, $\beta_{\text{min}}=0.5$, and $\beta_{\text{min}}=0.2$) first increases and then approaches a constant that is determined by $\eta(\beta_{\text {min}},\alpha)$ given in \eqref{loss} and has also been  shown in Table I. This is due to the fact that when $N$ is moderate, the signal power of the AP-user link is comparable to that of the IRS-user link, thus the power loss due to the IRS hardware imperfection is more pronounced with increasing $N$. However, when $N$ is sufficiently large such that the reflected signal power by the IRS dominates in the total receive power at the user, the power loss arising from the imperfect IRS reflection converges to that in accordance with the asymptotic analysis given in Proposition \ref{qm_pro}.
	
	Next, by varying  $d$, the required AP transmit power is compared in Fig. \ref{variable_d_power}  for the following schemes with  $N=40$: 
	\begin{enumerate} 
		\item Lower bound: solve (P1) with $\beta_{\text{min}}=1$ (i.e., ideal IRS) by using semidefinite relaxation (SDR) with Gaussian randomization which has been shown to achieve near-optimal performance in \cite{wu2018intelligent}.
		\item AO method (Algorithm 1): solve (P2) with $\beta_{\text{min}}=0.2$ (i.e., practical IRS) by using Proposition \ref{sub_allocation_pro} with the trust region  given by Proposition \ref{tr_pro}. 
		\item AO method (exhaustive search): solve (P2) with $\beta_{\text{min}}=0.2$  by using the 1D search.
		\item Penalty-based method (Algorithm 2):  solve (P3) with $\beta_{\text{min}}=0.2$. We set $\epsilon_1=10^{-3}$, $\epsilon_2=10^{-8}$, $\varrho=1.3$, $\mu^{(1)}=10^{-15}$, and $\Delta=0.05$.
		\item Ideal IRS assumption: the phase shifts designed for the IRS with $\beta_{\text{min}}=1$ are applied to the practical  IRS with $\beta_{\text{min}}=0.2$. 
		\item Upper bound: the system without using the  IRS by setting $ \mathbf w^* =\sqrt{P}\frac{\mathbf h_d}{\|\mathbf h_d\|}$ with
		$P^\star = \frac{\gamma \sigma^2}{\| \mathbf h_d\|^2}$.
		 
	\end{enumerate}
	Note that the initial phase shift values of the proposed penalty-based and AO algorithms, i.e., $\{\theta_n\}_{n=1}^N$, are randomly selected from $\{\pi,-\pi\}$ such that each reflecting element has the maximum reflection amplitude. 
	
	\begin{figure}
		\centering \vspace{-0.8cm}
		\begin{minipage}{.5\textwidth}
			\begin{subfigure}[b]{0.48\textwidth} 
				\centering
				\includegraphics[trim = 0mm 0mm 0mm 0mm, clip,width=\textwidth]{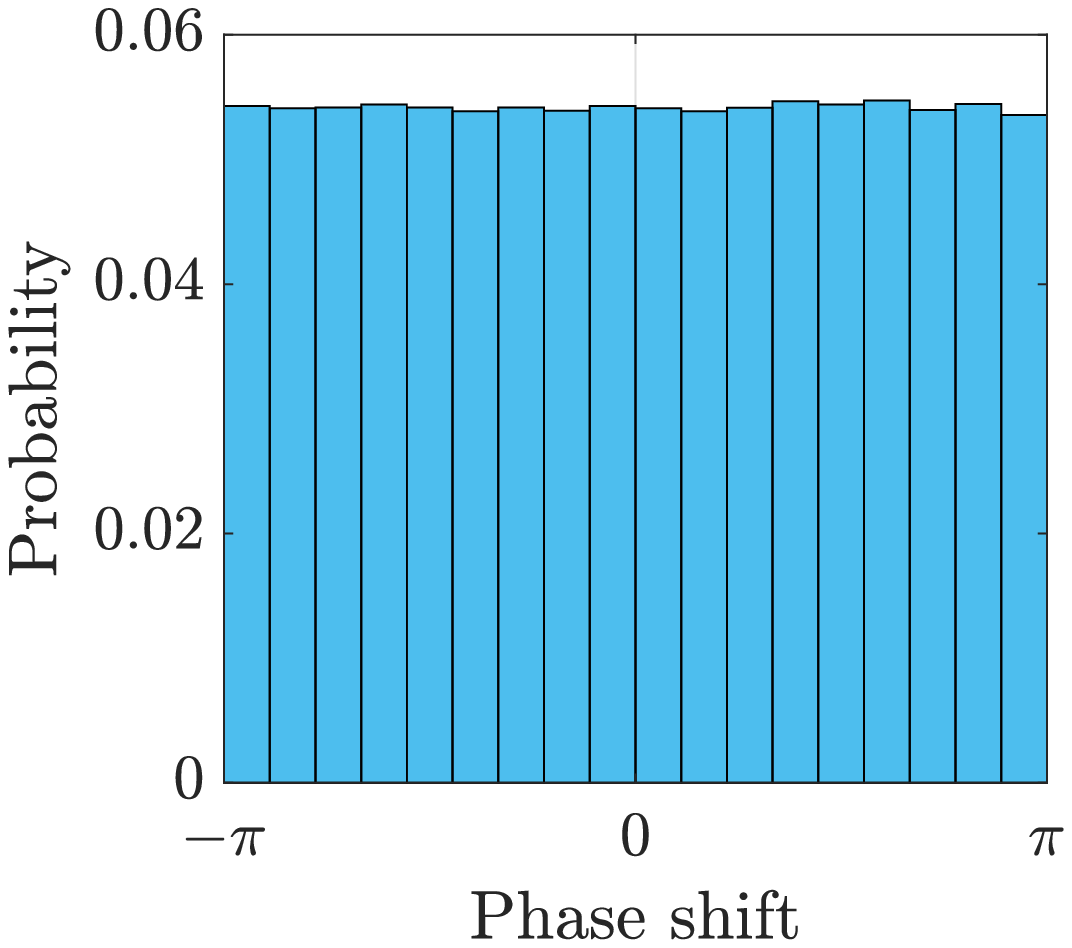}
				\caption{The Ideal IRS ($\beta_{\mathrm {min}}=1$)}
				\label{ideal_histo} 
			\end{subfigure}
			\hspace{0.1cm}
			\begin{subfigure}[b]{0.48\textwidth} 
				\centering
				\includegraphics[trim = 0mm 0mm 0mm 0mm, clip,width=\textwidth]{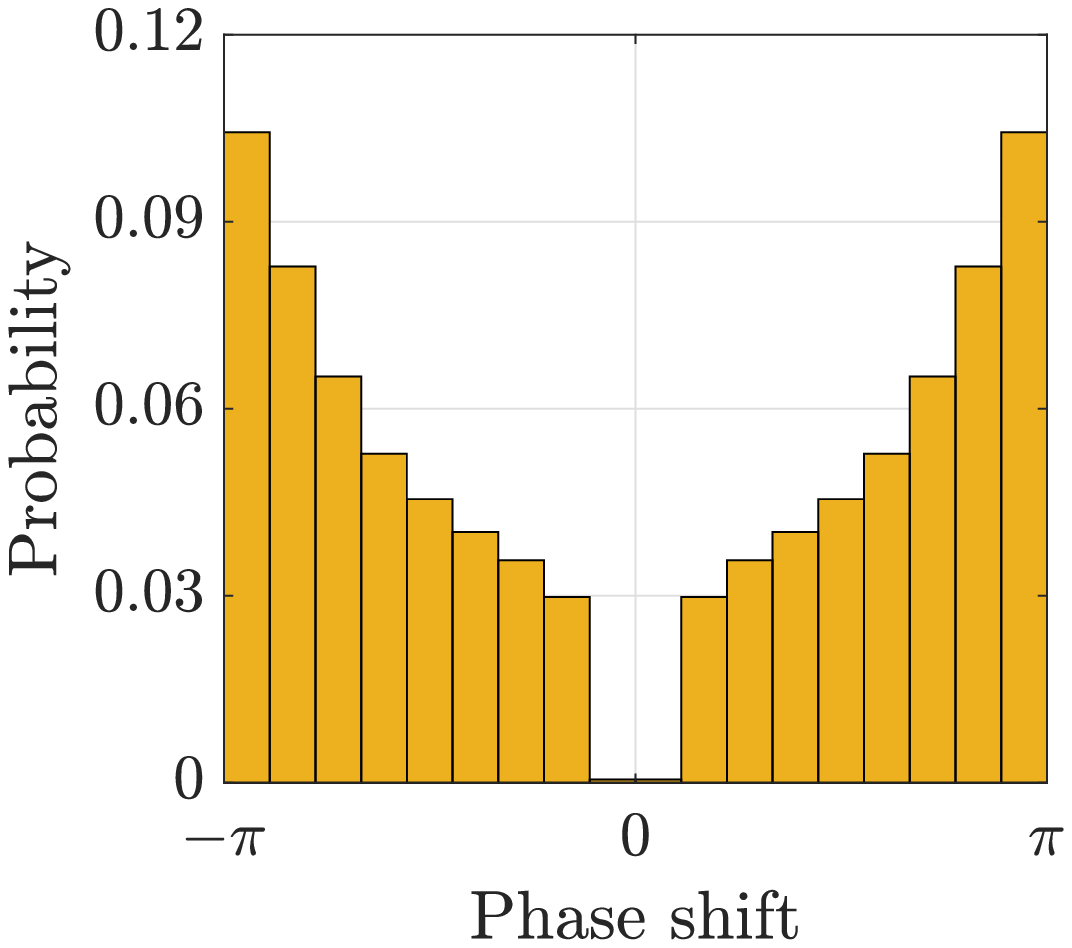}
				\caption{Practical IRS ($\beta_{\mathrm {min}}=0.2$)}
				\label{practical_histo} 
			\end{subfigure}
			\caption{Distribution of  IRS phase shift values over reflecting elements.} \label{phase_histo} 
		\end{minipage}
		\hspace{1cm}
		\begin{minipage}{.42\textwidth}
			\includegraphics[width=\textwidth,height=\textheight,keepaspectratio]{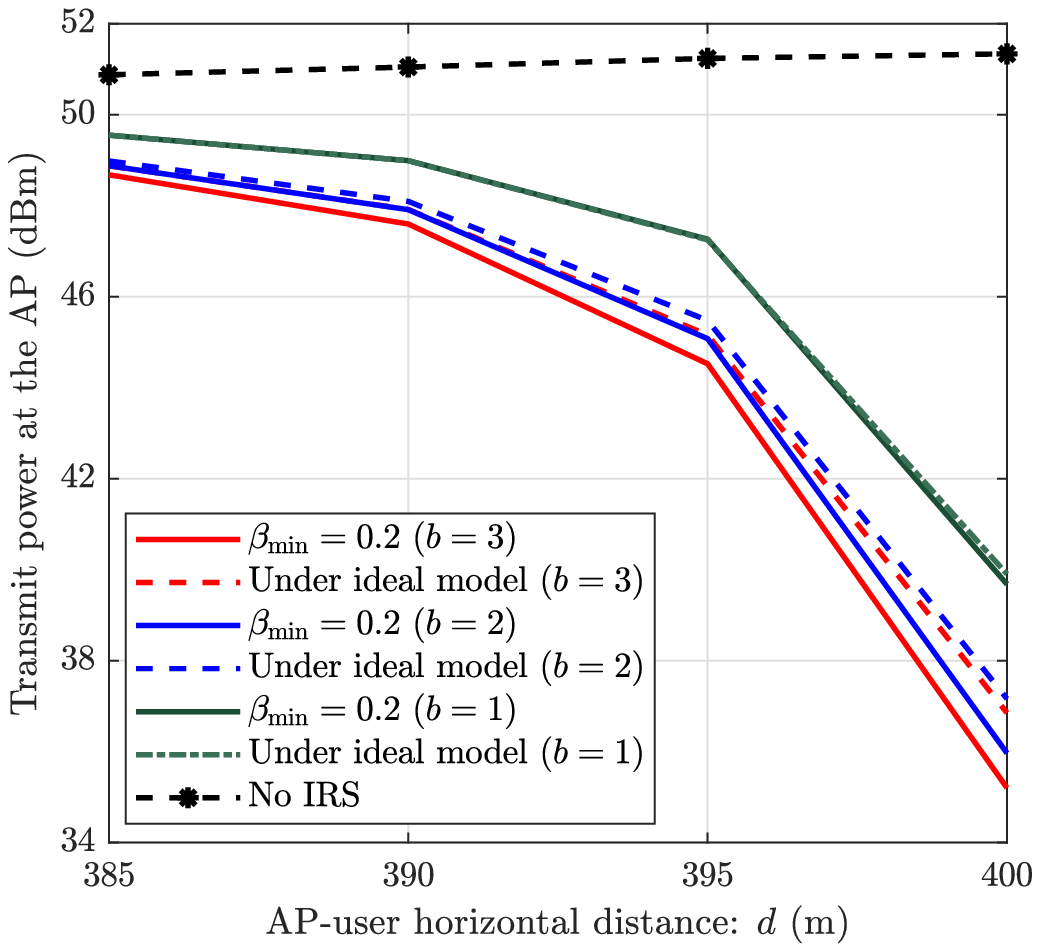}
			\caption{Achievable rate versus $d$ in the case of discrete phase shift.}
			\label{variable_d_d} 
		\end{minipage} \vspace{-0.8cm}
	\end{figure}
	
	It is observed from Fig. \ref{variable_d_power} that schemes  2), 3), and 4) perform very close to each other, and  significantly outperform schemes  5) and 6). This suggests that AO and penalty-based algorithms with Proposition \ref{sub_allocation_pro}  provide a practically appealing solution to (P1) considering their  low complexity. It is also observed that when the user moves closer to the IRS, the performance gap between the proposed schemes  and scheme 5) increases. This is due to the fact that the user benefits from the stronger reflecting channel via IRS ($\mathbf h_r$), and therefore proper   reflection design based on the practical  IRS model becomes more  crucial. In contrast, when the user moves toward  the AP, the above performance gap  decreases as the  AP-user direct channel ($\mathbf h_d$) becomes dominant, thus reducing  the effectiveness of the IRS reflection.
	
	Next, in Fig. \ref{phase_histo}, we plot the phase shift distribution over the  IRS reflecting elements. In particular, we solve (P1) by using  Algorithm 1 for the ideal IRS with  $\beta_{\text{min}}=1$ and the practical IRS with  $\beta_{\text{min}}=0.2$ to empirically obtain the phase shift distribution. It is observed from Fig. \ref{phase_histo}(a) that when $\beta_{\text{min}}=1$, the phase shift value is uniformly distributed in  $[-\pi,  \pi)$. However, when $\beta_{\text{min}}=0.2$, the probability increases from zero to $\pi$ or $-\pi$ while there is nearly  zero probability near the zero phase shift, as shown in  Fig. \ref{phase_histo}(b). This is expected since the minimum reflection amplitude (i.e., $\beta_{\text{min}}$) occurs  at zero  phase shift and asymptotically approaches the maximum of one at  $\pi$ or $-\pi$, thus  the  optimized phase shift  is more concentrated towards either  $\pi$ or $-\pi$ to maximize the effective channel power gain.

	In practice, it is  difficult to implement  continuous phase shifts at each of the reflecting elements \cite{IRS_discrete}. To take this into account,  we consider the practical setup where  the phase shift at each element of the IRS can only take a finite number of discrete values, which are assumed to be  equally spaced in $[-\pi,\pi)$. Denote by $b$ the number of bits used to represent each of the levels. Then the set of phase shifts at each element is given by $\mathcal F = \{0,\Delta \theta,\dots,\Delta \theta(U-1)\}$ where $\Delta \theta = 2\pi/U$ and $U=2^b$.  In Fig. \ref{variable_d_d}, we compare the AP transmit power for different values of $b$ when the user moves closer to the IRS with $\beta_{\mathrm {min}}=0.2$ and $N=40$. By  solving (P2) with 1D search over $\mathcal F$, we compare the performance of the following two schemes for designing the discrete IRS phase shifts: i) based on the actual  value of $\beta_{\mathrm {min}}$ (i.e., $0.2$); and ii) based on the ideal model (i.e., assuming  $\beta_{\mathrm {min}}=1$).
	It is observed that for $b=1$, both schemes have nearly  the same performance and thus the consideration of IRS hardware imperfection is not necessary. However, when $b$ increases, the performance gap between these  two schemes increases. This is expected as $b$ increases,   the reflected signal power by the IRS is more dominant  in the total receive power at the user, thus the performance loss due to the inaccurate (ideal) phase shift model  becomes more pronounced.
	 
	\subsection{Multiuser Case } \label{simulation2}
	
	Next, we consider a multiuser system with $K = 4$, $M=4$, $N=40$, $d_x=3.5$ m,  $d_y=400$ m, and $r=2.5$ m. Without loss of generality, we assume that all users have the same SINR target, i.e., $\gamma_k=\gamma,\forall k$.  We set $\epsilon_1=10^{-3}$, $\epsilon_2=10^{-5}$, $\varrho=1.3$, $\mu^{(1)}=10^{-14}$, $\nu^{(1)}=10$, and $\Delta=0.05$.  Other system 	parameters are the same as in Section \ref{simulation1} (if not specified otherwise).

\begin{figure}
		\centering \vspace{-0.8cm}
		\begin{minipage}{.49\textwidth} 
			\includegraphics[width=\textwidth,height=\textheight,keepaspectratio]{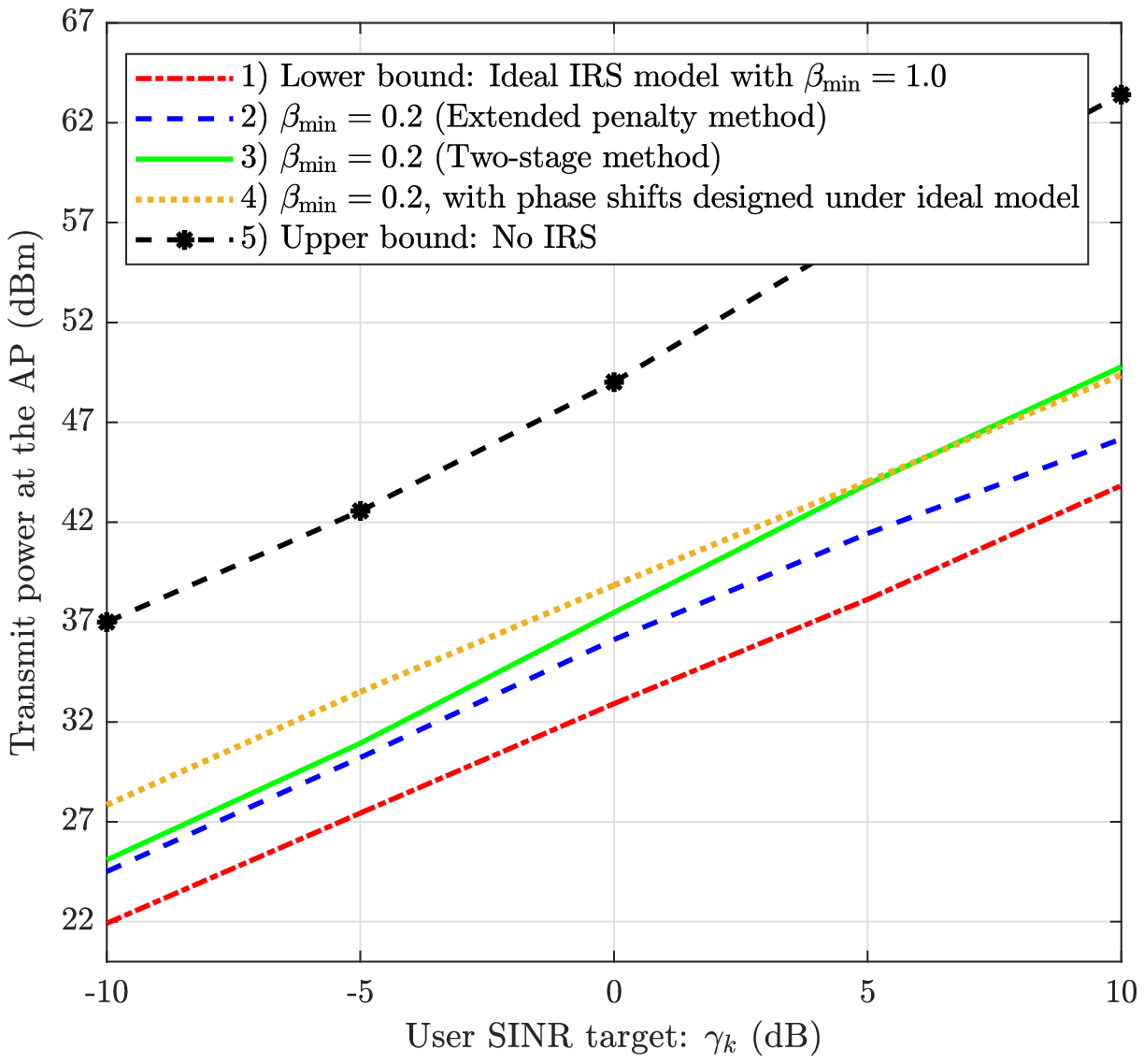}
			\caption{AP transmit power versus user SINR target.}  
			\label{variable_sinr} 
		\end{minipage}
		\hspace{0.0cm}
		\begin{minipage}{.49\textwidth} 
			\includegraphics[width=\textwidth,height=\textheight,keepaspectratio]{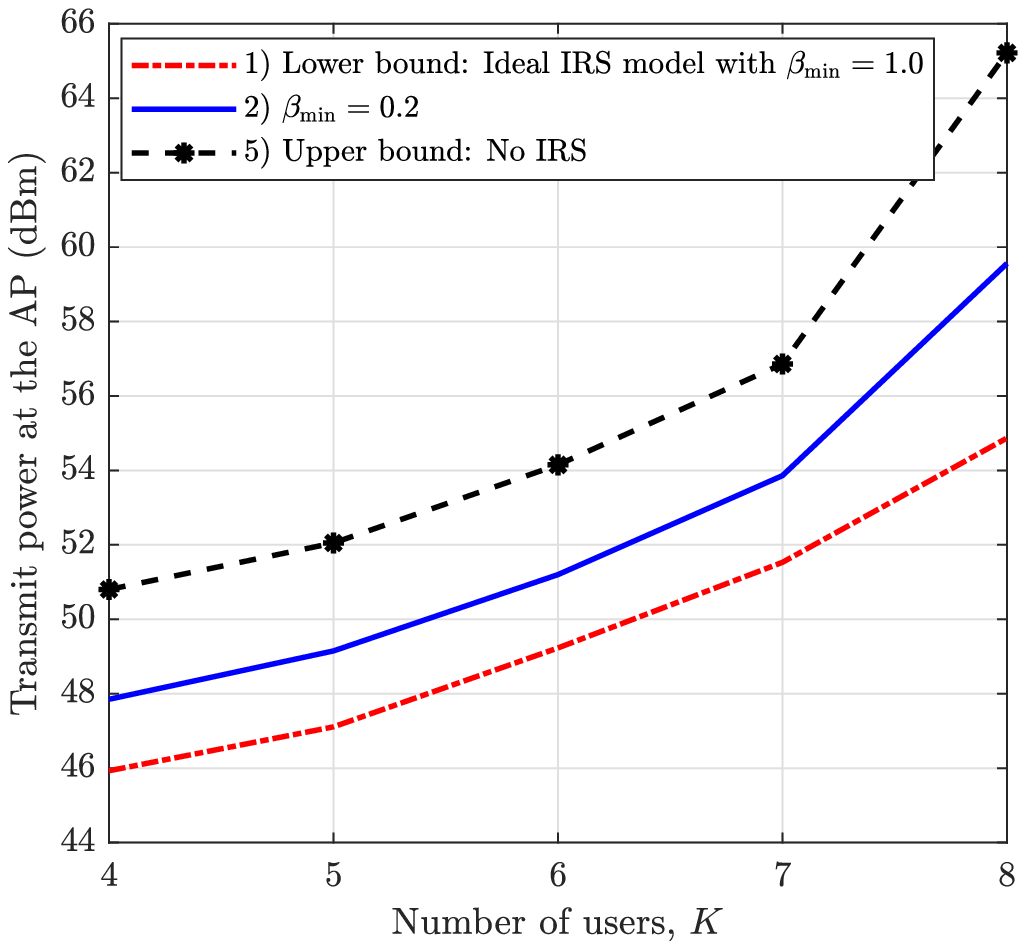}
			\caption{AP transmit power versus number of users.}
			\label{variable_k_mu_0} 
		\end{minipage} \vspace{-0.2cm}
	\end{figure}

	By fixing  $d = 400$ m and varying the SINR target, in Fig. \ref{variable_sinr},  we plot the required AP transmit power for the following schemes: 
	\begin{enumerate} 
		\item Lower bound: solve (P6) with $\beta_{\text{min}}=1$   using the extended penalty-based method.
		\item Solve (P6) with $\beta_{\text{min}}=0.2$  using the extended penalty-based method.
		\item Solve (P7.1) and (P7.1) with $\beta_{\text{min}}=0.2$  using the two-stage method.
		\item Ideal IRS assumption: phase shifts obtained by  the extended penalty-based algorithm by assuming $\beta_{\text{min}}=1$, but applied to a practical  IRS with $\beta_{\text{min}}=0.2$.
		\item Upper bound: the system without using an IRS by solving (P7.2) with $\mathbf h_k =  \mathbf h_{d,k} $. 
	\end{enumerate}
	It is observed that the performance gap between the extended  penalty-based method and two-stage method  increases as $\gamma$ increases. Although the two-stage method suffers from small performance loss in the low SINR regime compared to the extended  penalty-based method, it performs even worse than scheme 4) in the high SINR regime. This is because the multiuser interference becomes the performance bottleneck  when the user SINR target is high, and thus the joint transmit beamforming and  IRS reflect beamforming design becomes more crucial, which requires the use of both the  practical phase shift model as well as more sophisticated optimization.  
	
	In Fig. \ref{variable_k_mu_0}, we show the AP transmit power
	versus the number of users, $K$,  by setting $\gamma_k=10$ dB, $M = 8$,  and $d=395$ m. In particular, we successively add  users to the cluster  and obtain  the required AP transmit power  using the extended-penalty-based method. It  is  observed that the required  transmit power increases  from $K=4$ to $K=7$ while the  performance gap between  different schemes remains almost constant. However, when the number of users becomes equal to  that of antennas at the AP, i.e., $K=M=8$, the AP transmit power in the case without IRS increases more drastically than that with the ideal IRS assuming  $\beta_{\mathrm {min}}=1$. This is due to the fact that adding the $8$-th user results in a poorly-conditioned MIMO channel in the case without IRS, while adding  the IRS with $\beta_{\mathrm {min}}=1$ is able to transform the overall effective  MIMO channel to be well-conditioned by adding strong multi-paths via IRS reflection. However, by applying the practical IRS with  $\beta_{\mathrm {min}}=0.2$, the IRS reflection is not strong enough to recover the full
	spatial multiplexing gain  and thus results in considerable  power loss. 
	
	\section{Conclusion} \label{con_sec}
	
	In this paper, we proposed the  practical IRS phase shift model and validated its accuracy based on experimental results. Under  the new model, we formulated and solved   joint transmit and reflect beamforming optimization problems in  IRS-aided multiuser systems to minimize the  transmit power at the AP subject to users' individual SINR constraints, by applying the AO and penalty-based optimization techniques. 
	Our simulation results  showed  that beamforming optimization  based on the conventional ideal phase shift model, which has been  widely used in the literature, may lead to significant performance loss as compared to the proposed practical model for both single-user and multiuser setups. In future work, it is worth investigating such performance difference in more general IRS-aided wireless communication systems, such as  OFDM-based system, NOMA-based system, physical layer security system, SWIPT system, and so on.

	
	\section*{Appendix A: Proof of Proposition \ref{qm_pro}} \label{single_QM}
	Since the IRS-reflected signal dominates in the user's receive signal power for asymptotically large $N$, the signal  from the AP-user link ($\mathbf h_{d}$) can be ignored. Thus, $P_{\mathrm{practical}}$ is approximately given by 
	$P_{\mathrm{practical}} \approx P\mathbb{E} \Big(\|(\mathbf v^H  \diag(\mathbf h_{r}^H) \mathbf G )\|^2\Big),$ with $\mathbf v(n) = \beta_n(\theta_n) e^{\jmath \theta_n} $, $\beta_n(\theta_n) =  (1-\beta_{\text {min}}) \left( \frac{\sin(\theta_n - \phi) +1}{2} \right)^k + \beta_{\text {min}}$. Let $\pmb \theta=[\theta_1,\dots,\theta_N]$. When $M=1$ and assuming the ideal phase shift model, i.e.,  $\beta_n(\theta_n)=1,\forall n$ in \eqref{eq:P1_C1}, the optimal solution of 	$\pmb \theta$ to (P1) is given by $\theta_n^*=-\arg( { {}} \mathbf G(n,1))+\arg( { {\mathbf h_r(n)}} ), \forall n$ \cite{wu2018intelligent}. Then we have  
	\begin{align} 
	\|(\mathbf v^H  \diag(\mathbf h_{r}^H) \mathbf G )\|^2 &= \Bigg| \Big(\sum_{n=1}^{N} |\mathbf v(n)||\mathbf h_r(n)||\mathbf G(n,1)|\Big)  \Bigg|^2\nonumber 
	=   \Bigg(\sum_{n=1}^{N} |\mathbf v(n)|^2|\mathbf h_r(n)|^2|\mathbf G(n,1)|^2 \nonumber \\ 
	&\hspace{15mm} {}  + \sum_{n=1}^{N} \sum_{i\neq n}^{N} |\mathbf v(n)||\mathbf h_r(n)||\mathbf G(n,1)| |\mathbf v(i)||\mathbf h_r(i)||\mathbf G(i,1)| \Bigg).  
	\end{align} 
	Note that $|\mathbf h_r(n)|$ and $|\mathbf G(n,m)|$ follow Rayleigh distribution with mean values $\sqrt{\pi}\varrho_r/2 $ and $\sqrt{\pi}\varrho_g/2 $. 	
	Since $|\mathbf v(n)|$, $|\mathbf h_r(n)|$, and $|\mathbf G(n,m)|$  are statistically independent, we have
	\begin{align} 
	\mathbb{E} \bigg(|\mathbf v(n)|^2|\mathbf h_r(n)|^2|\mathbf G(n,1)|^2 \bigg) &= \varrho_r^2 \varrho_g^2 \mathbb{E}\big(|\mathbf v(n)|^2\big) = \varrho_r^2 \varrho_g^2 \mathbb{E}\big(\beta_n(\theta_n)^2\big), \nonumber \\
	\mathbb{E}\bigg(|\mathbf v(n)||\mathbf h_r(n)||\mathbf G(n,1)|\bigg) &= \frac{\pi \varrho_r \varrho_g \mathbb{E}\big(|\mathbf v(n)|\big)}{4}= \frac{\pi \varrho_r \varrho_g \mathbb{E}\big(\beta_n(\theta_n)\big)}{4}. \nonumber
	\end{align}
	It then follows that
	\begin{align} 
	P_{\mathrm{practical}} &=  P\mathbb{E} \Big(\|(\mathbf v^H  \diag(\mathbf h_{r}^H) \mathbf G )\|^2\Big) \nonumber
	= PN \varrho_r^2 \varrho_g^2 \mathbb{E}\big(\beta_n(\theta_n)^2\big) + PN(N-1) \frac{\pi^2 \varrho_r^2 \varrho_g^2 \mathbb{E}\big(\beta_n(\theta_n)\big)^2}{16}. 
	\end{align} 
	On the other hand, under the same phase shift solution $\pmb \theta$, for  the ideal IRS model with the unity amplitude at each reflecting  element regardless of the phase shift, i.e.,  
	 $\beta_n(\theta_n)=1,\forall n$, we have
	\begin{align} 
	P_{\mathrm{ideal}}
	&= PN \varrho_r^2 \varrho_g^2  + PN(N-1) \frac{\pi^2 \varrho_r^2 \varrho_g^2 }{16}, 
	\end{align}
	since  $\mathbb{E}\big(\beta_n(\theta_n)^2\big)=1$ and $\mathbb{E}\big(\beta_n(\theta_n)\big)=1$. As a
	result, when $N\rightarrow \infty$, the ratio between $P_{\mathrm{practical}}$ and $P_{\mathrm{ideal}}$ is given by
		\begin{equation} 
	\eta(\beta_{\text {min}},k)=\frac{P_{\mathrm{practical}}}{P_{\mathrm{ideal}}} = \mathbb{E}\big(\beta_n(\theta_n)\big)^2=\Bigg( \frac{1}{2\pi}\int_{-\pi}^{\pi} \beta(\theta) \ \ d\theta  \Bigg)^2,
	\end{equation}
	with  $\beta(\theta) =(1-\beta_{\text {min}}) \left( \frac{\sin(\theta - \phi) +1}{2} \right)^k + \beta_{\text {min}}$, since $ \theta$ is uniformly distributed in $[-\pi, \pi]$.  This thus  completes the proof.

	\section*{Appendix B: Proof of Proposition \ref{tr_pro}} \label{single_ER_1}
	
	Let $\delta \geq0$ be a sufficiently small constant.
	First, considering the case of $\arg(\varphi_n) \geq 0$ or  $\lambda =0$, the following inequalities are obtained:
		\begin{itemize} 
		\item Since $\beta_n(\arg(\varphi_n)) \geq \beta_n(\arg(\varphi_n)-\delta)$ and $\cos(\delta)\leq1$, $f(\arg(\varphi_n)) \geq f(\arg(\varphi_n)-\delta)$.
		\item Since $\beta_n(\pi) \geq \beta_n(\pi+\delta)$ and $\cos(\delta)\leq1$, $f(\pi) \geq f(\pi+\delta)$.
	\end{itemize} 
	It then follows that $\exists \delta \in [0,\pi-\arg(\varphi_n)]$ such that
	$f(\arg(\varphi_n)+\delta) \geq f(\arg(\varphi_n))$ and thus  $\theta_n^* \in [\arg(\varphi_n),\pi]$. Similarly, we can show   $\theta_n^* \in [\arg(\varphi_n),-\pi]$ when $\arg(\varphi_n) < 0$ or $\lambda =1$. The proof  is thus completed.
	 	
	\section*{Appendix C: Proof of Proposition \ref{sub_allocation_pro}}  \label{single_ER_2}
	Given three  points $\theta_A$, $\theta_B$,  $\theta_C$ and their corresponding function values $f_1$, $f_2$, $f_3$, we seek to determine three constants $a_0$, $a_1$, and $a_2$ such that the following  quadratic function is constructed,  
	\begin{align} \label{q_f}
	g(\theta_n) = a_0 + a_1(\theta_n-\theta_A) + a_2(\theta_n-\theta_A)(\theta_n-\theta_B).
	\end{align}
	When $\theta_n=\theta_A$, $\theta_n=\theta_B$, and $\theta_n=\theta_C$, the constants $a_0$, $a_1$, and $a_2$ can be respectively obtained. Substituting them into the stationary point of $g(\theta_n)$, i.e., $\hat \theta_n^\star = \frac{\theta_A+\theta_B}{2} - \frac{a_1}{2a_2}$, allows us to obtain \eqref{our_method}. The proof  is thus completed.

	\section*{Appendix D: Proof of Proposition \ref{tr_pro2}} \label{single_ER_3}
	Let $\Delta >0$,  $f(\theta_n) \defeq 2\beta_n(\theta_n)|v_n|\cos(\psi_n-\theta_n) - \beta_n^2(\theta_n)$,  and $\lambda =0$ when $\psi_n \geq 0$  or $\lambda =1$ otherwise. If $f(\psi_n-\Delta) > f(\psi_n)$ for $\lambda =0$, we have
	\begin{align} 
	\frac{\beta_n^2(\psi_n) - \beta_n^2(\psi_n-\Delta)   }{2(\beta_n(\psi_n) - \beta_n(\psi_n-\Delta)\cos(\Delta))} > |v_n|. \label{xy}
	\end{align}
	Since $\beta_n(\psi_n) - \beta_n(\psi_n-\Delta)\cos(\Delta) > \beta_n(\psi_n) - \beta_n(\psi_n-\Delta)$, \eqref{xy} can be simplified as
	\begin{align} 
	\frac{\beta_n(\psi_n) + \beta_n(\psi_n-\Delta)   }{2} > |v_n|. \label{xyz}
	\end{align}
	Likewise, if $f(\psi_n+\Delta) > f(\psi_n)$ for $\lambda =0$, the  following inequality can be obtained similarly  by following the above  steps. 
	\begin{align} 
	\frac{\beta_n(\psi_n) + \beta_n(\psi_n+\Delta)   }{2} < |v_n|. \label{xyzz}
	\end{align}
	It is not difficult to observe that always $\exists \Delta$ such that either \eqref{xyz} or \eqref{xyzz} holds depending on the values of  $v_n$ and $\psi_n$. Following the similar  steps as above allows us to obtain  \eqref{pro3eq} for the case of $\lambda =1$. The proof  is thus completed. 
	
	\bibliographystyle{IEEEtran}  
	\footnotesize{\bibliography{bibfile}}

\begin{thebibliography}{10}
\providecommand{\url}[1]{#1}
\csname url@samestyle\endcsname
\providecommand{\newblock}{\relax}
\providecommand{\bibinfo}[2]{#2}
\providecommand{\BIBentrySTDinterwordspacing}{\spaceskip=0pt\relax}
\providecommand{\BIBentryALTinterwordstretchfactor}{4}
\providecommand{\BIBentryALTinterwordspacing}{\spaceskip=\fontdimen2\font plus
\BIBentryALTinterwordstretchfactor\fontdimen3\font minus
  \fontdimen4\font\relax}
\providecommand{\BIBforeignlanguage}[2]{{%
\expandafter\ifx\csname l@#1\endcsname\relax
\typeout{** WARNING: IEEEtran.bst: No hyphenation pattern has been}%
\typeout{** loaded for the language `#1'. Using the pattern for}%
\typeout{** the default language instead.}%
\else
\language=\csname l@#1\endcsname
\fi
#2}}
\providecommand{\BIBdecl}{\relax}
\BIBdecl

\bibitem{my_icc}
S.~Abeywickrama, R.~Zhang, and C.~Yuen, ``Intelligent reflecting surface:
  {Practical} phase shift model and beamforming optimization,'' [Online].
  Available: https://arxiv.org/abs/1907.06002.

\bibitem{wu2018intelligent}
Q.~{Wu} and R.~{Zhang}, ``Intelligent reflecting surface enhanced wireless
  network via joint active and passive beamforming,'' \emph{IEEE Trans.
  Wireless Commun.}, vol.~18, no.~11, pp. 5394--5409, Nov. 2019.

\bibitem{qq_magazine}
Q.~Wu and R.~Zhang, ``Towards smart and reconfigurable environment:
  {Intelligent} reflecting surface aided wireless network,'' \emph{IEEE Commun.
  Mag.}, vol.~58, no.~1, pp. 106--112, Jan. 2020.

\bibitem{chongwang}
C.~{Huang}, A.~{Zappone}, G.~C. {Alexandropoulos}, M.~{Debbah}, and C.~{Yuen},
  ``Reconfigurable intelligent surfaces for energy efficiency in wireless
  communication,'' \emph{IEEE Trans. Wireless Commun.}, vol.~18, no.~8, pp.
  4157--4170, Aug. 2019.

\bibitem{8796365}
E.~{Basar}, M.~{Di Renzo}, J.~{De Rosny}, M.~{Debbah}, M.~{Alouini}, and
  R.~{Zhang}, ``Wireless communications through reconfigurable intelligent
  surfaces,'' \emph{IEEE Access}, vol.~7, pp. 116\,753--116\,773, Aug. 2019.

\bibitem{chongwen_hollo}
C.~Huang~\textit{et al.}, ``Holographic {MIMO} surfaces for {6G} wireless
  networks: {Opportunities}, challenges, and trends,'' [Online]. Available:
  https://arxiv.org/abs/1911.12296.

\bibitem{8723525}
M.~{Cui}, G.~{Zhang}, and R.~{Zhang}, ``Secure wireless communication via
  intelligent reflecting surface,'' \emph{IEEE Wireless Commun. Lett.}, vol.~8,
  no.~5, pp. 1410--1414, Oct. 2019.

\bibitem{8972400}
X.~{Guan}, Q.~{Wu}, and R.~{Zhang}, ``Intelligent reflecting surface assisted
  secrecy communication: {Is} artificial noise helpful or not?'' \emph{IEEE
  Wireless Commun. Lett., DOI:10.1109/LWC.2020.2969629}, Jan. 2020.

\bibitem{xu2019resource}
D.~Xu, X.~Yu, Y.~Sun, D.~W.~K. Ng, and R.~Schober, ``Resource allocation for
  secure {IRS}-assisted multiuser {MISO} systems,'' [Online]. Available:
  https://arxiv.org/abs/1907.03085.

\bibitem{irs_ofdm}
Y.~Yang, B.~Zheng, S.~Zhang, and R.~Zhang, ``Intelligent reflecting surface
  meets {OFDM}: Protocol design and rate maximization,'' [Online]. Available:
  https://arxiv.org/abs/1906.09956.

\bibitem{IRS_channel_estimation}
B.~{Zheng} and R.~{Zhang}, ``Intelligent reflecting surface-enhanced {OFDM}:
  Channel estimation and reflection optimization,'' \emph{IEEE Wireless Commun.
  Lett., DOI:10.1109/LWC.2019.2961357}, Dec. 2019.

\bibitem{yang2019intelligent}
G.~Yang, X.~Xu, and Y.-C. Liang, ``Intelligent reflecting surface assisted
  non-orthogonal multiple access,'' [Online]. Available:
  https://arxiv.org/abs/1907.03133.

\bibitem{8970580}
B.~{Zheng}, Q.~{Wu}, and R.~{Zhang}, ``Intelligent reflecting surface-assisted
  multiple access with user pairing: {NOMA} or {OMA}?'' \emph{IEEE Commun.
  Lett., DOI:10.1109/LCOMM.2020.2969870}, Jan. 2020.

\bibitem{8941080}
Q.~{Wu} and R.~{Zhang}, ``Weighted sum power maximization for intelligent
  reflecting surface aided {SWIPT},'' \emph{IEEE Wireless Commun. Lett.,
  DOI:10.1109/LWC.2019.2961656}, Dec. 2019.

\bibitem{IRS_SWIPT_2}
Q.~Wu and R.~Zhang, ``Joint active and passive beamforming optimization for
  intelligent reflecting surface assisted {SWIPT} under {QoS} constraints,''
  [Online]. Available: https://arxiv.org/abs/1910.06220.

\bibitem{nadeem2019intelligent}
Q.-U.-A. Nadeem, A.~Kammoun, A.~Chaaban, M.~Debbah, and M.-S. Alouini,
  ``Intelligent reflecting surface assisted multi-user {MISO} communication,''
  [Online]. Available: https://arxiv.org/abs/1906.02360.

\bibitem{wang2019joint}
P.~Wang, J.~Fang, and H.~Li, ``Joint beamforming for intelligent reflecting
  surface-assisted millimeter wave communications,'' [Online]. Available:
  https://arxiv.org/abs/1910.08541.

\bibitem{zhang2019capacity}
S.~Zhang and R.~Zhang, ``Capacity characterization for intelligent reflecting
  surface aided {MIMO} communication,'' [Online]. Available:
  https://arxiv.org/abs/1910.01573.

\bibitem{pan2019multicell}
C.~Pan, H.~Ren, K.~Wang, W.~Xu, M.~Elkashlan, A.~Nallanathan, and L.~Hanzo,
  ``Multicell {MIMO} communications relying on intelligent reflecting
  surface,'' [Online]. Available: https://arxiv.org/abs/1907.10864.

\bibitem{fu2019intelligent}
M.~Fu, Y.~Zhou, and Y.~Shi, ``Intelligent reflecting surface for downlink
  non-orthogonal multiple access networks,'' [Online]. Available:
  https://arxiv.org/abs/1906.09434.

\bibitem{mu2019exploiting}
X.~Mu, Y.~Liu, L.~Guo, J.~Lin, and N.~Al-Dhahir, ``Exploiting intelligent
  reflecting surfaces in multi-antenna aided {NOMA} systems,'' [Online].
  Available: https://arxiv.org/abs/1910.13636.

\bibitem{zhao2019intelligent}
M.-M. Zhao, Q.~Wu, M.-J. Zhao, and R.~Zhang, ``Intelligent reflecting surface
  enhanced wireless network: Two-timescale beamforming optimization,''
  [Online]. Available: https://arxiv.org/abs/1912.01818.

\bibitem{4619755}
H.~{Rajagopalan} and Y.~{Rahmat-Samii}, ``Loss quantification for microstrip
  reflectarray: Issue of high fields and currents,'' in \emph{Proc. IEEE
  Antennas and Propag. Society Int. Symposium}, Jul. 2008, pp. 1--4.

\bibitem{phase_dependent_amplitude}
W.~Tang~\textit{et al.}, ``{MIMO} transmission through reconfigurable
  intelligent surface: {System} design, analysis, and implementation,''
  [Online]. Available: https://arxiv.org/abs/1912.09955.

\bibitem{zhu2013active}
B.~O. Zhu, J.~Zhao, and Y.~Feng, ``Active impedance metasurface with full 360
  reflection phase tuning,'' \emph{Scientific reports}, vol.~3, pp. 3059--3064,
  Oct. 2013.

\bibitem{irs_amp}
M.~E. {Bialkowski}, A.~W. {Robinson}, and H.~J. {Song}, ``Design, development,
  and testing of {X}-band amplifying reflectarrays,'' \emph{IEEE Trans. on
  Antennas and Propag.}, vol.~50, no.~8, pp. 1065--1076, Aug. 2002.

\bibitem{tang2018wireless}
W.~Tang, X.~Li, J.~Y. Dai, S.~Jin, Y.~Zeng, Q.~Cheng, and T.~J. Cui, ``Wireless
  communications with programmable metasurface: Transceiver design and
  experimental results,'' [Online]. Available:
  https://arxiv.org/abs/1811.08119.

\bibitem{PhysRevApplied}
F.~Liu~\textit{et al.}, ``Intelligent metasurfaces with continuously tunable
  local surface impedance for multiple reconfigurable functions,'' \emph{Phys.
  Rev. Applied}, vol.~11, no.~4, pp. 44\,024--44\,033, Apr. 2019.

\bibitem{koziel2013surrogate}
S.~Koziel and L.~Leifsson, \emph{Surrogate-based modeling and
  optimization}.\hskip 1em plus 0.5em minus 0.4em\relax Springer, 2013.

\bibitem{microwave_book}
D.~M. Pozar, \emph{Microwave Engineering (3th Edition)}.\hskip 1em plus 0.5em
  minus 0.4em\relax {N}ew {Y}ork {J}ohn {W}iley {\&} {S}ons, 2005.

\bibitem{powell}
R.~P. Brent, \emph{Algorithms for Minimization without Derivatives}.\hskip 1em
  plus 0.5em minus 0.4em\relax Prentice-Hall, 1973.

\bibitem{penalty_method}
Q.~{Shi}, M.~{Hong}, X.~{Gao}, E.~{Song}, Y.~{Cai}, and W.~{Xu}, ``Joint
  source-relay design for full-duplex {MIMO} {AF} relay systems,'' \emph{IEEE
  Trans. Signal Process.}, vol.~64, no.~23, pp. 6118--6131, Dec. 2016.

\bibitem{mm_tsp}
Y.~{Sun}, P.~{Babu}, and D.~P. {Palomar}, ``Majorization-minimization
  algorithms in signal processing, communications, and machine learning,''
  \emph{IEEE Trans. Signal Process.}, vol.~65, no.~3, pp. 794--816, Feb. 2017.

\bibitem{1262126}
M.~{Schubert} and H.~{Boche}, ``Solution of the multiuser downlink beamforming
  problem with individual {SINR} constraints,'' \emph{IEEE Trans. Veh.
  Technol.}, vol.~53, no.~1, pp. 18--28, Jan. 2004.

\bibitem{1664998}
{Zhi-Quan Luo} and {Wei Yu}, ``An introduction to convex optimization for
  communications and signal processing,'' \emph{IEEE J. Sel. Areas Commun.},
  vol.~24, no.~8, pp. 1426--1438, Aug. 2006.

\bibitem{IRS_discrete}
Q.~{Wu} and R.~{Zhang}, ``Beamforming optimization for wireless network aided
  by intelligent reflecting surface with discrete phase shifts,'' \emph{IEEE
  Trans. Commun., DOI:10.1109/TCOMM.2019.2958916}, Dec. 2019.

\end{thebibliography}
\end{document}